%% file: MAINARDI_M-L-4REV_ARXIV-FABRIZIO.tex
\documentclass[12pt]{article}

\usepackage{graphicx}

\def\e{\hbox{e}}

\def\ds{\displaystyle}
\def\RR{\vbox {\hbox to 8.9pt {I\hskip-2.1pt R\hfil}}}
\def\NN{{\rm I\hskip-2pt N}}
\def\CC{{\rm C\hskip-4.8pt \vrule height 6pt width 12000sp\hskip 5pt}}
\def\vsp{\par} 
\def\q{\quad}  
\def\cen{\centerline}
\def \rec#1{{\frac{1}{#1}}}
\def\pni{\par \noindent}
\def\vsh{\vskip 0.25truecm\noindent}

\def\vsp{\vsh\pni}
\begin{document}

\cen{{\bf FRACALMO PRE-PRINT: \ http://www.fracalmo.org}}
\vsh
\hrule
\vskip 0.50truecm
\font\title=cmbx12 scaled\magstep2
\font\bfs=cmbx12 scaled\magstep1
\font\little=cmr10
\begin{center}
{\bfs On some properties of  the Mittag-Leffler function $\mathbf{E_\alpha(-t^\alpha)}$,} 
\\
{\bfs completely monotone for $\mathbf{t> 0}$  with  $\mathbf{0<\alpha<1}$}
\end{center}
\begin{center}
{\bf Francesco MAINARDI}
\\
{Department of Physics, University of Bologna, and INFN}
 \\{Via Irnerio 46,  Bologna, I-40126 ITALY}
   \\{{\tt francesco.mainardi@unibo.it} ;  {\tt francesco.mainardi@bo.infn.it}} 
\end{center}



\bigskip
\noindent
{\bf This paper has been published in   
Discrete and Continuous Dynamical Systems - Series B (DCDS-B),
Vol. 19, No 7 (2014),
pp. 2267-2278. DOI: 10.3934/dcdsb.2014.19.2267,
 in the special issue devoted to Prof. Mauro Fabrizio, University of Bologna.}

\begin{abstract}
We analyse some peculiar properties of the function  
of the Mittag-Leffler (M-L) type, 
$e_\alpha(t) := E_\alpha(-t^\alpha)$  for $0<\alpha<1$ and $t>0$, which  is known to 
 be completely monotone (CM)  with a non-negative spectrum of frequencies and times,
suitable to model fractional relaxation processes.
 We first note that (surprisingly)  these two spectra coincide so providing 
a universal scaling  property of  this function, not  well pointed out in the literature.
Furthermore, we consider  the problem of approximating our  M-L  function  with simpler CM  functions for small and large times. 
We provide two different sets of elementary CM functions that are asymptotically  equivalent to $e_\alpha(t)$
as $t\to 0$ and $t\to +\infty$.
The first set is given by the stretched exponential for small times 
and the power law for large times, following a standard approach.    
For the second set we chose  two rational CM functions in $t^\alpha$,  
obtained as the  Pad\`e  Approximants  (PA)  $[0/1]$  to the convergent series in positive powers  (as $t\to 0$) 
and to the asymptotic series in negative powers  (as $t\to \infty$),  respectively.  
From numerical computations we are allowed to the conjecture that the second set provides upper and lower bounds to the Mittag-Leffler function. 
\end{abstract}
\section{Introduction}
\vsp
Since  a few decades the special transcendental function  known as Mittag-Leffler function has attracted 
an increasing  attention of  researchers  because of its key role in treating problems related to integral and differential equations of fractional order. 
\vsp 
Since its introduction by the Swedish mathematician Mittag-Leffler at the beginning of the last century up to the 1990's,  this function was seldom considered by  mathematicians and applied scientists.
\vsp
Before the 1990's, from a mathematical point of view ,we recall the
1930  paper by Hille and Tamarkin \cite{Hille-Tamarkin_1930} 
on the solutions of the Abel integral equation of the second kind, 
and the  books by Davis \cite{Davis_BOOK1936},
Sansone \&  Gerretsen \cite{Sansone-Gerretsen_BOOK1960},
Dzherbashyan \cite{Dzherbashyan_BOOK1966} (unfortunately in Russian),
and finally  Samko et al. \cite{SKM_BOOK1993}.
Particular mention would be for
 the 1955 Handbook of  High Transcendental Functions of the Bateman project \cite{Erdelyi_HTF1955}, 
 where this function  was  treated in the chapter devoted to miscellaneous functions.
For former applications we recall 
an interesting  note by Davis \cite{Davis_BOOK1936} reporting a previous research   by Dr. Kenneth S. Cole  in connection with nerve conduction, 
and the  papers    by  Cole \& Cole \cite{Cole-Cole_1942}, 
Gross \cite{Gross_JAP1947}  and  
Caputo \& Mainardi 
\cite{Caputo-Mainardi_PAGEOPH1971,Caputo-Mainardi_RNC1971},
where the Mittag-Leffler function was adopted to represent 
the responses in dielectric and   viscoelastic media.
\vsp
In the 1960's  the Mittag-Leffler function started to exit from the realm of miscellaneous functions because it was considered as a special case of the general  class of  Fox $H$ functions, that 
can exhibit an arbitrary number of  parameters in their integral Mellin-Barnes representation, see e.g.the books by 
Kiryakova \cite{Kiryakova_BOOK1994}, 
 Kilbas and Saigo  \cite{Kilbas-Saigo_BOOK2004}, 
Marichev  \cite{Marichev_BOOK1983},
 Mathai \& Saxena \cite{Mathai-Saxena_BOOK1978},  
 Mathai et al. \cite{Mathai-Saxena-Haubold_BOOK2010}, 
 Srivastava et.al. \cite{Srivastava-Gupta-Goyal_BOOK1982}.
\vsp
However, in our opinion, this classification in  a too general framework has,  to some extent, obscured the relevance and the applicability of this function in applied science. 
In fact    most mathematical models are based  on a small  number of parameters, say  1 or 2 or 3, 
so that a general theory may be confusing whereas the adoption of a generalized Mittag-Leffler function 
with 2 or 3 indices may be sufficient, see e.g.
Beghin \& Orsingher \cite{Beghin-Orsingher_EPJ2010}, 
Capelas et al. \cite {Capelas-et-al_EPJ-ST2011},
  Sandev et al. \cite{Sandev_FLE_FCAA2012},
 Tomovski et al. \cite{Tomovski-et-al_ITSF2010}. 
Multi-index Mittag-Leffler functions have been introduced as well,
see e.g.   
Kilbas et al. \cite{Kilbas-et-al_FCAA2013}, 
Kilbas \& Saigo \cite{Kilbas-Saigo_1995},
Kiryakova \cite{Kiryakova_CMA2010}, 
 Kiryakova \&  Luchko \cite{Kiryakova-Luchko_2010}, 
 but their extensive use has  not yet been  pointed out in applied sciences
  up to now.  
\vsp   
Nowadays it is well recognized that the Mittag-Leffler function   plays  a fundamental role  in  Fractional Calculus   
even if with a single parameter (as originally introduced by Mittag-Leffler)  just to be worth of being  referred to as  the {\it Queen Function of Fractional Calculus}, see Mainardi \&  Gorenflo \cite{Mainardi-Gorenflo_FCAA2007}.
On this respect  we  just point out  some  
 reviews and treatises on Fractional Calculus  (in order of publication time): 
  Gorenflo \& Mainardi \cite{Gorenflo-Mainardi_CISM1997},
  Podlubny \cite{Podlubny_BOOK1999},  
  Hilfer \cite{Hilfer_BOOK2000}, 
  Kilbas et al. \cite{Kilbas-Srivastava-Trujillo_BOOK2006},  
  Magin \cite{Magin_BOOK2006}, 
  Mathai \& Haubold \cite{Mathai-Haubold_BOOK2008}, 
  Mainardi \cite{Mainardi_BOOK2010}, 
  Diethelm \cite{Diethelm_BOOK2010},  
  Tarasov \cite{Tarasov_BOOK2011},
   Klafter et al. \cite{KLM_BOOK2012}, 
   Baleanu et al. \cite{BDST_BOOK2012},
   Uchaikin \cite{Uchaikin_BOOK2013}.
\section{The Mittag-Leffler function in fractional \\ relaxation processes}
\noindent
The Mittag-Leffler function is defined by the following power series,
convergent in the whole complex plane,
$$
E_\alpha (z) := \sum_{n=0}^\infty \frac{z^n}{\Gamma (\alpha n+1)}
\,,\q \alpha > 0\,, \q z\in \CC\,. \eqno(2.1)$$
We recognize that it is an entire function   providing  a simple generalization of the
  exponential function to which it reduces for $\alpha=1$.
 We also note that for the convergence of the power series in (2.1)
 the parameter $\alpha$ may be complex provided that $\Re (\alpha) >0$.
The most interesting properties of the Mittag-Leffler function
are associated with its asymptotic expansions
as $z \to \infty$ in various sectors of the complex plane.
For  detailed asymptotic analysis,  
which includes the smooth transition across the Stokes lines,  
 the interested reader is referred to Wong and Zhao \cite{Wong-Zhao_CA2002}.
\vsp
In this paper we limit ourselves to the Mittag-Leffler function of order $\alpha \in (0,1)$ on the negative real semi-axis where is known to be completely monotone (CM) due a classical  result by  Pollard \cite{Pollard_BAMS1948}, see also 
Feller \cite{Feller_BOOK1971}.
\vsp
Let us recall that a function $ \phi(t)$  with $t\in\RR^+$ is    called a  completely
monotone (CM) function if it is non-negative, of class $C^{\infty}$,  and
$(-1)^n \phi^{(n)}(t)\ge 0$ for all $n \in \NN$. 
Then a function $ \psi(t)$  with $t\in\RR^+$ is    called a Bernstein function  
if it is non-negative, of class $C^{\infty}$, with a CM first derivative.
These functions play fundamental roles in linear hereditary mechanics to represent relaxation and creep processes, see e.g Mainardi \cite{Mainardi_BOOK2010}. 
For mathematical details we refer the interested reader  to the survey paper 
by Miller and  Samko \cite{Miller-Samko_ITSF2001}
and to the recent book by Schilling et al. \cite{SSV_BOOK2012}.  
\vsp   
In particular we are interested to the function 
$$ e_\alpha(t) := E_\alpha(-t^\alpha)
 =    
\sum_{n=0}^\infty (-1)^n \frac{t^{\alpha n}}{\Gamma (\alpha n+1)}
\,, \quad t>0\,, \quad 0<\alpha \le 1\,,\eqno (2.2)$$
that provides the solution to the fractional relaxation equation, 
see Gorenflo and Mainardi \cite{Gorenflo-Mainardi_CISM1997},
Mainardi and  Gorenflo \cite{Mainardi-Gorenflo_FCAA2007}, 
Mainardi \cite{Mainardi_BOOK2010}.
\vsp
For readers'  convenience let us briefly outline the topic 
concerning the   generalization  via fractional calculus 
 of the first-order differential equation governing the
 phenomenon of (exponential) relaxation.  
Recalling   (in non-dimensional units)
 the initial value problem 
 $$\frac{du}{dt} = -u(t) \,, \quad t\ge 0\,, \quad \hbox{with}\quad u(0^+)= 1\,\eqno(2.3)$$
 whose solution is
 $$u(t) = \exp (-t)\,,\eqno (2.4)$$
  the following  two alternatives 
  with $\alpha \in (0,1)$ are offered in the literature:
  $$\frac{du}{dt } = - D_t^{1-\alpha}\,u(t) \,,\quad  t\ge 0\,, \quad \hbox{with}\quad u(0^+)= 1\,,\eqno(2.5a)$$
  $$ _*D_t^\alpha \,u(t) = -u(t)\,, \quad t\ge 0\,, \quad \hbox{with}\quad u(0^+)= 1\,.\eqno(2.5b)$$
where  
 $D_t^{1-\alpha}$ and $\,  _*D_t^\alpha$ denote the fractional derivative of order $1-\alpha$ in 
 the Riemann-Liouville sense and the fractional derivative of order $\alpha$ in the Caputo sense, respectively.
 \vsp
 For a generic order $\mu\in (0,1)$ and for a sufficiently well-behaved function $f(t)$ with $t\in \RR^+$ 
 the above derivatives are defined as follows, see e.g. 
 Gorenflo and Mainardi \cite{Gorenflo-Mainardi_CISM1997}, 
 Podlubny \cite{Podlubny_BOOK1999}, 
$$ D_t^\mu  \,f(t) =
  {\ds  \rec{\Gamma(1-\mu )}}\, 
{\ds \frac{d}{dt}}\left[  
  \int_0^t
    \! \frac{f(\tau)}{ (t-\tau )^{\mu }}\,d\tau\right]\,, \eqno(2.6a)  $$ 
$$  _*D_t^\mu  \,f(t) =
{\ds \rec{\Gamma(1-\mu )}}\int_0^t
    \! \frac{f^\prime (\tau)}{ (t-\tau )^{\mu }}\,d\tau\,. \eqno(2.6b)  $$   
  Between the two derivatives we have the  relationship 
   $$
	 {\ds _*D_t^\mu  \,f(t)} = 
  {\ds D_t^\mu \,f(t) -  f(0^+)\, \frac{t^{-\mu}} {\Gamma(1-\mu )}}  =
 D_t^\mu  \,\left[ f(t) - f(0^+) \right] \, .\eqno(2.7)
   $$
Both derivatives in the limit $\mu \to 1^-$ reduce to the standard first derivative but for 
$\mu \to 0^+$ we have 
$$ D_t^\mu f(t) \to f(t)\,, \quad  _*D_t^\mu f(t) = f(t) - f(0^+)\,, \quad \mu \to 0^+\,.\eqno(2.8)$$  
 \vsp
 In analogy to the standard problem (2.3), we  solve the problems (2.5a) and (2.5b)
  with the Laplace transform technique,  using  the rules pertinent to the corresponding fractional derivatives.
 The problems (a) and (b) are equivalent since  the Laplace transform of the solution
 in both cases comes out as
 $$  \widetilde u(s) = \frac{s^{\alpha-1}}{s^\alpha +1}\,, \eqno(2.9) $$
that yields our function
$$ u(t) = e_\alpha(t) := E_\alpha(-t^\alpha)\,.\eqno(2.10)$$
The Laplace transform pair
$$       e_\alpha(t) \,\div \, \frac{s^{\alpha-1}}{s^\alpha +1}\,, \quad \alpha >0 \,, \eqno(2.11)$$
 can be proved   by transforming term by term the power series representation  of $e_\alpha(t)$
 in the R.H.S of  (2.2).
 Furthermore,    by
 anti-transforming the R.H.S  of (2.11)  by using the complex Bromwich formula, and taking into account  for $0<\alpha<1$ the contribution from  branch cut on the negative real semi-axis (the denominator $s^\alpha +1$  does nowhere vanish in the cut plane $-\pi \le \hbox{arg} \, s \le \pi $),  we get, 
 see also Gorenflo and Mainardi \cite{Gorenflo-Mainardi_CISM1997},
$$ e_\alpha(t) = \int_0^\infty \!\!  \e^{-rt} K_\alpha(r)\, dr\,, \eqno(2.12) $$
where     
$$      K_\alpha(r) = 
 \mp \,\rec{\pi}\,    {\rm Im}\,
  \left\{ \left. \frac{s^{\alpha -1}} {s^\alpha +1}\right\vert_{{\ds s=r\, \e^{\pm i\pi}}} \right\}
 = \rec{\pi}\,
   \frac{ r^{\alpha -1}\, \sin \,(\alpha \pi)}
    {r^{2\alpha} + 2\, r^{\alpha} \, \cos \, (\alpha \pi) +1} \ge 0\,.
      \eqno(2.13)$$
    Since   $K_\alpha(r)$ is non-negative for all $r$ in the integral, the above formula proves 
    that $e_\alpha(t)$ is  CM function in view of the Bernstein theorem. This theorem provides a necessary and sufficient condition for a CM function as a real Laplace transform of a non-negative measure. 
\vsp
However, the CM property of $e_\alpha(t)$  can also be seen  as a consequence  of the result by  Pollard
because the transformation $x=t^\alpha$ is a Bernstein function for $\alpha \in (0,1)$.
In fact it is known that a CM function can be obtained by composing a CM with a Bernstein function based on the following theorem:
{\it Let $\phi(t)$ be a CM function and let $\psi(t)$ be a Bernstein function, then $\phi[\psi(t)]$  is a CM function.} 
 \vsp
  As a matter of fact,  $K_\alpha(r)$ provides an interesting  spectral representation of  $e_\alpha(t)$ in frequencies. With the change of variable $\tau=1/r$ we get the corresponding spectral representation in relaxation times, namely
$$ e_\alpha(t) = \int_0^\infty \e^{-t/\tau} H_\alpha(\tau)\, d\tau\,, \;
H_\alpha(\tau) = \tau^{-2} \, K_\alpha(1/\tau)\,, \eqno(2.14) $$
that can be interpreted as a continuous distribution of elementary (i.e. exponential) relaxation processes.
As a consequence    we get the identity between  the two spectral distributions, that is 
$$ H_\alpha(\tau)
= \rec{\pi}\,
   \frac{ \tau^{\alpha -1}\, \sin \,(\alpha \pi)}
    {\tau^{2\alpha} + 2\, \tau^{\alpha} \, \cos \, (\alpha \pi) +1} 
\,, \eqno(2.15)$$
a surprising fact pointed out in Linear Viscoelasticity 
by the author in his  book \cite{Mainardi_BOOK2010}.
This kind of universal/scaling  property seems a peculiar one for our Mittag-Leffler function $e_\alpha(t)$.
In Fig 1 we show $K_\alpha(r)$ for some values of the parameter $\alpha$.
 Of course for $\alpha=1$ the Mittag-Leffler function reduces to the exponential function $\exp(-t)$
 and the  corresponding spectral distribution is the    Dirac delta generalized function centred at $r=1$, namely $\delta(r-1)$.
\vsp
 \begin{center}
\includegraphics[width=7.5cm]{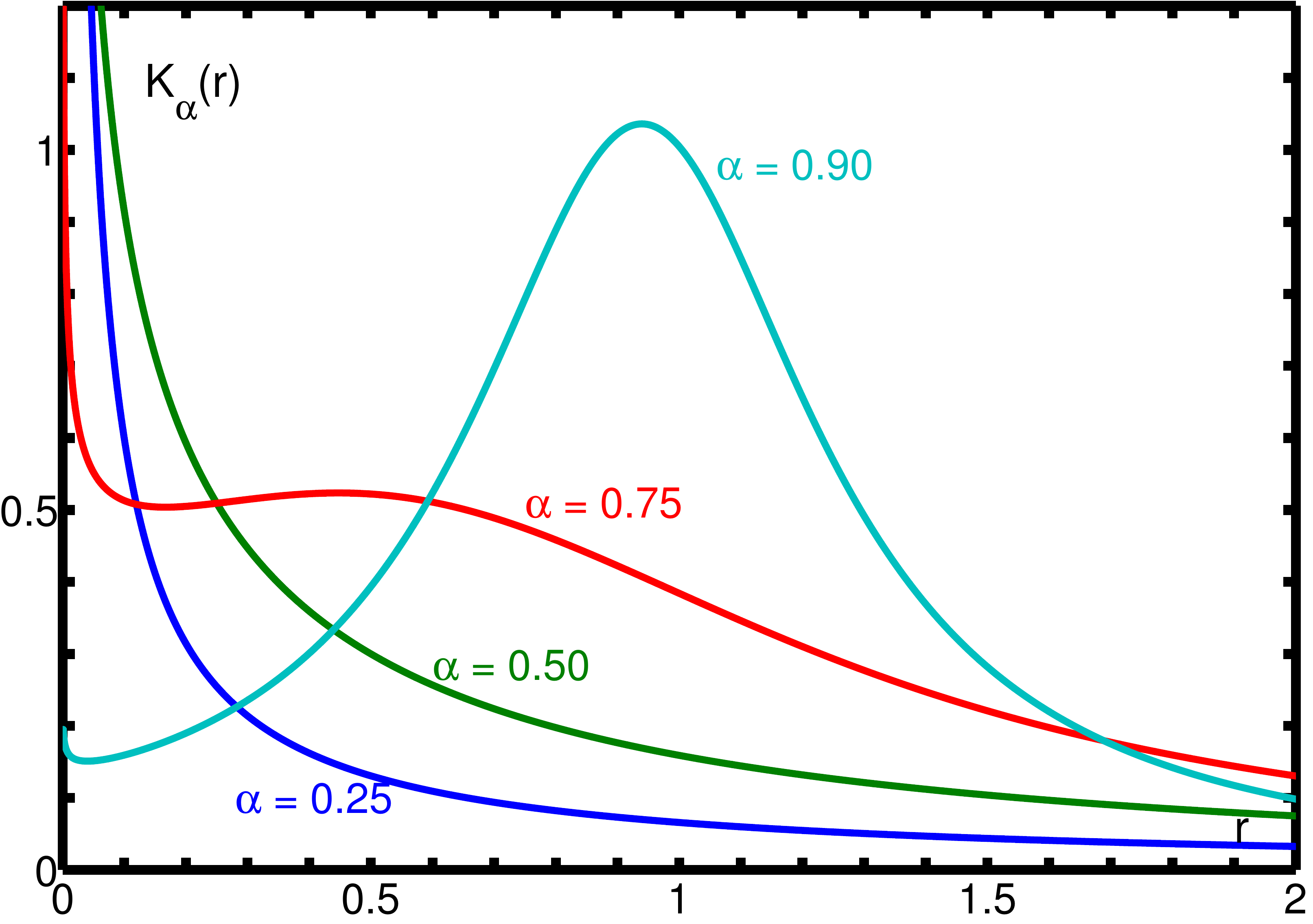}
\end{center}
{{\bf Fig.1} Plots of the spectral function  $K_\alpha(r) $ for $\alpha=0.25, 0.50, 0.75, 0.90$ in the frequency range  $0\le r \le 2$.}
\section{Asymptotic approximations to the Mittag-Lefler function}
\vsp
In Fig 2  we show some plots of $e_\alpha(t)$  for some values of the parameter $\alpha$.
It is worth to note the different rates of decay of $e_\alpha(t) $   for small and large times.
In fact the decay is very fast as $t\to 0^+$ and very slow as $t \to +\infty$.
\vsp 
\begin{center}.
\includegraphics[width=7.5cm]{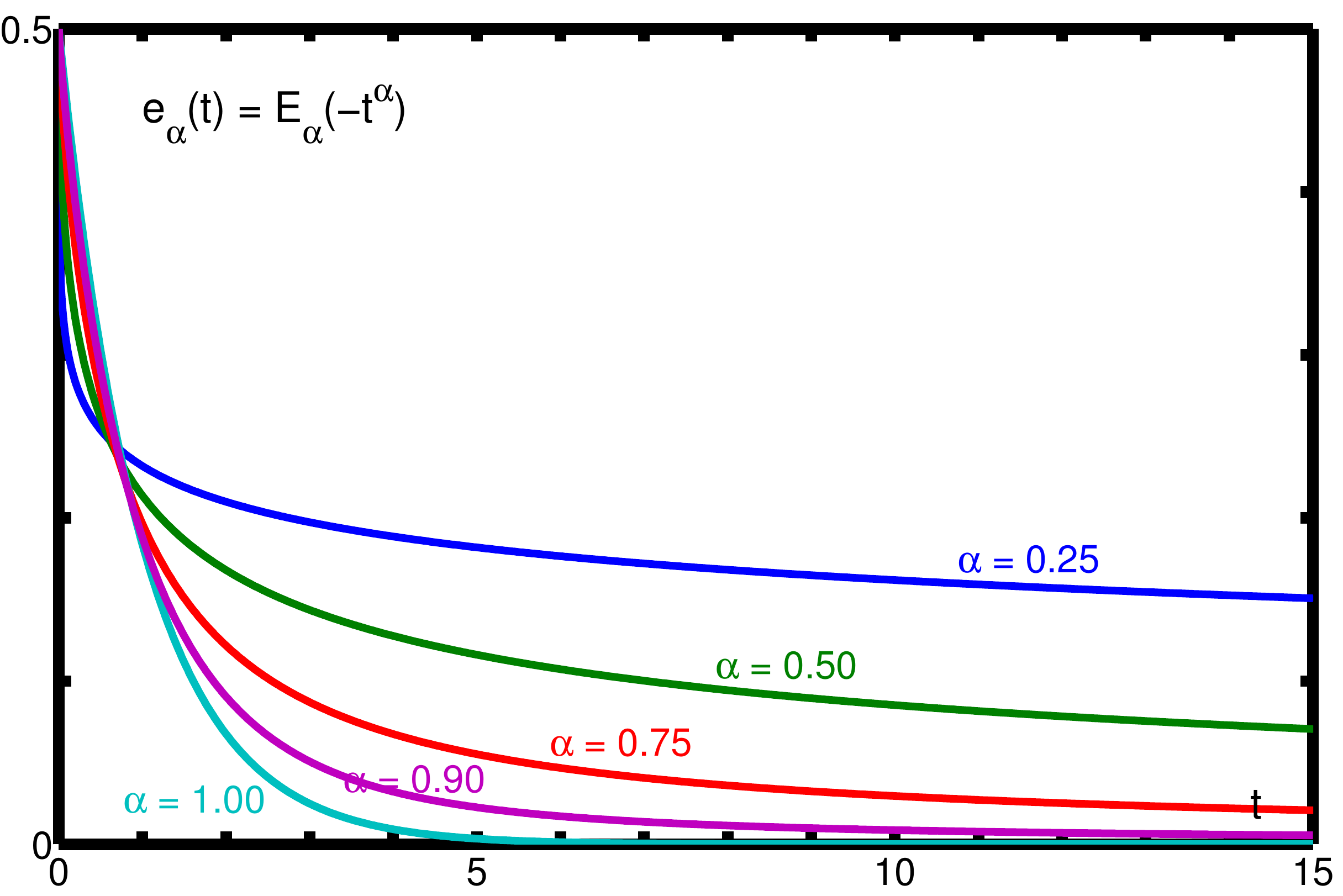} 
\end{center}
{{\bf Fig.2} Plots of the Mittag-Leffler function $e_\alpha(t)$  for $\alpha=0.25, 0.50, 0.75, 0.90, 1.$ in the time range $0\le t\le 15$ }
\vsp
\subsection{The two common asymptotic approximations}
  It is common to point out that     the function $e_\alpha(t)$  matches for $t\to 0^+$ with a stretched exponential    with an infinite negative derivative,   whereas as $t\to \infty$ with a negative power law.
The short time  approximation is  derived from the convergent power series representation (2.2). In  fact,
$$ e_\alpha(t) = 1 - \frac{t^\alpha}{\Gamma(1+\alpha)} + \dots 
\sim  \exp{\ds \left[- \frac{t^\alpha}{\Gamma(1+\alpha )}\right]}\,, \quad  t\to 0\, . \eqno(3.1)$$
The long time approximation is derived from the asymptotic power series representation of $e_\alpha(t)$ that turns out to be, see Erd\'elyi (1955),
 $$ e_\alpha(t) \sim 
 \sum_{n=1}^\infty  (-1)^{n-1} \,\frac{ t^{-\alpha n}}{\Gamma(1- \alpha n)}\,, 
 \quad t\to \infty\,,\eqno(3.2) $$
 so that, at the first order, 
 $$e_\alpha(t) \sim  {\ds \frac{t^{-\alpha}}{\Gamma(1-\alpha )} }\,, \quad t\to \infty\,. \eqno(3.3) $$
 As a consequence  the function $e_\alpha(t)$  interpolates 
 for intermediate time $t$ between the stretched exponential
and the negative power law.
The stretched exponential models   
  the  very fast decay  for small  time $t$, whereas the asymptotic   power law 
  is due to the very slow decay for large  time $t$.
In fact, we have the two commonly stated  asymptotic  representations: 
  $$
\e_\alpha  (t) \sim
\left\{ 
\begin{array}{ll}
 e_\alpha^0(t) := \exp{\ds \left[- \frac{t^\alpha}{\Gamma(1+\alpha )}\right]}\,, 
  &  t\to 0\,;   \\ \\
e_\alpha^\infty(t) := {\ds \frac{t^{-\alpha}}{\Gamma(1-\alpha )} } = 
{\ds \frac{\sin (\alpha  \pi)}{\pi}\,\frac{\Gamma(\alpha )}{t^\alpha }}\,,
 &  t\to \infty\,.
\end{array}
\right . \eqno(3.4)
   $$
   The stretched exponential   replaces  the rapidly decreasing expression   
    $1- {t^\alpha}/{\Gamma(1+\alpha )} $  from (3.1).
 Of course, {\it for sufficiently small and for sufficiently large values of $t$} 
  we have the inequality
   $$ e_\alpha^0(t) \le e_\alpha^\infty (t)\,,  \quad 0<\alpha<1\,. \eqno (3.5)$$
   \vsp
In  Figs  3-7 LEFT  we compare for $\alpha = 0.25, 0.5, 0.75, 0.90, 0.99$ in   logarithmic scales  the function $e_\alpha(t)$ (continuous line) and its asymptotic representations, the stretched exponential $e^0_\alpha(t)$ valid for $ t\to 0$ (dashed line)  and the power law $e^\infty_\alpha(t)$  valid for $t\to \infty$ (dotted line).
We have chosen the time range  $10^{-5} \le t \le 10^{+5}$.
In the RIGHT we have shown  the plots of the relative errors (in absolute values)
$$  \frac{|e^0_\alpha(t)- e_\alpha(t)|}{e_\alpha(t)}\,, \quad
\frac{|e^\infty_\alpha(t)- e_\alpha(t)|}{e_\alpha(t)}\,, \eqno (3.6)$$
pointing out a continuous line at an error 1\% under which the approximations can be considered reliable. 

\begin{center}
\includegraphics[width=6.35cm]{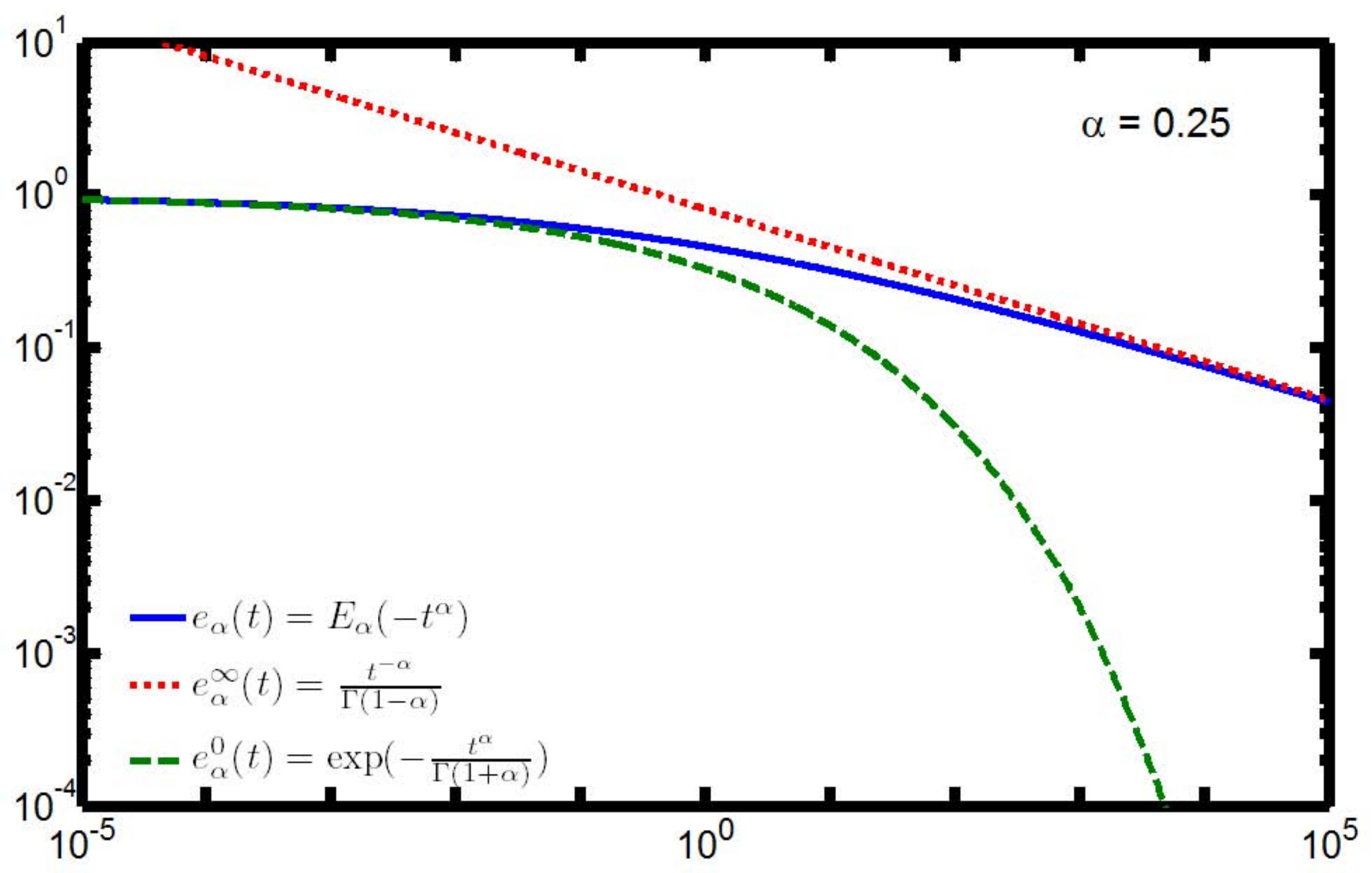}
\includegraphics[width=6.2cm]{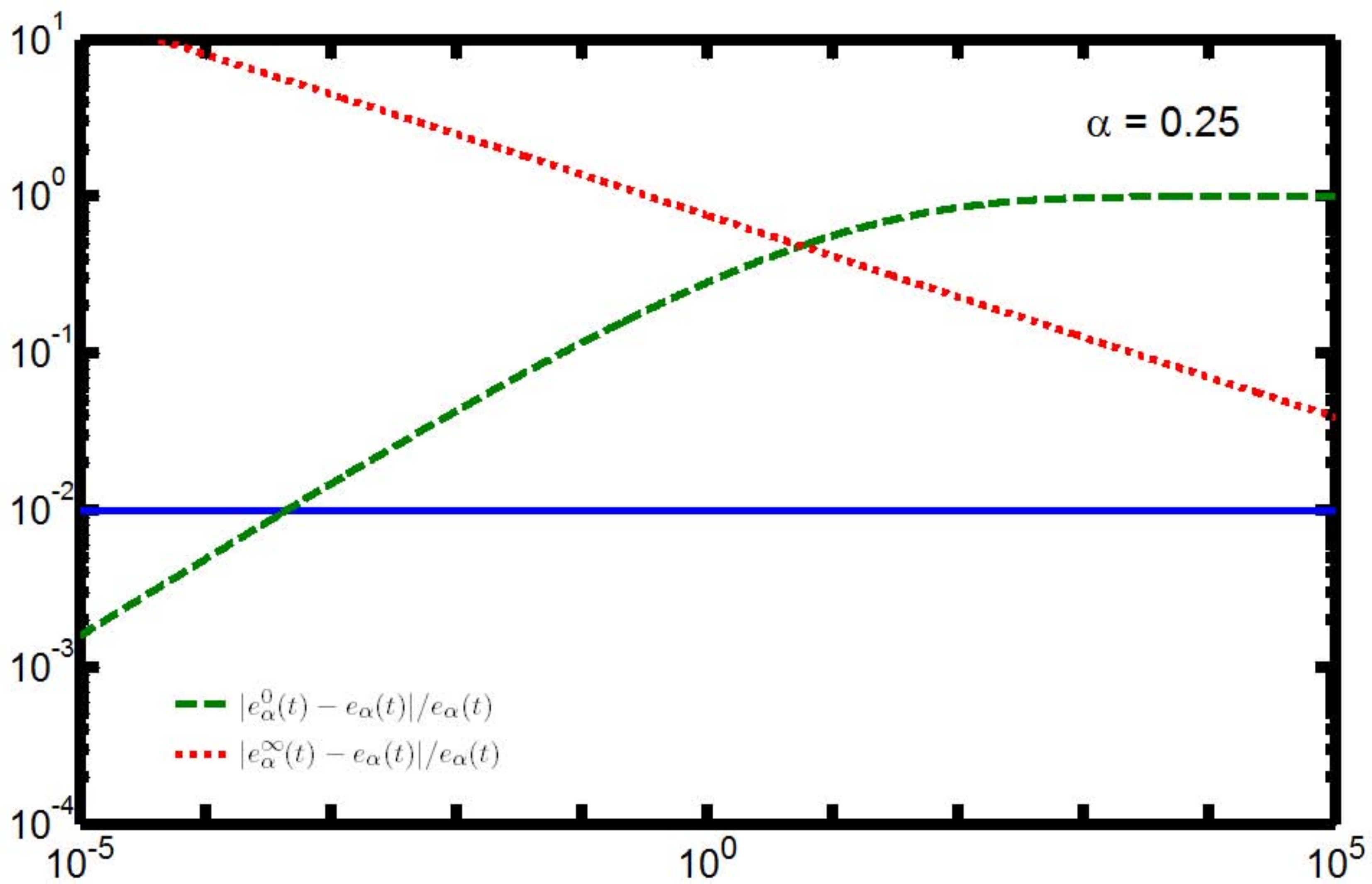}
\end{center}
{{\bf Fig.3} Approximations $e^0_\alpha(t)$ (dashed line) and $ e^\infty_\alpha(t)$ 
(dotted line) to   $e_\alpha(t)$  (LEFT) 
and  the corresponding relative errors (RIGHT)  
 in   $10^{-5} \le t \le 10^{+5}$ for $\alpha=0.25$.}
\smallskip
\begin{center}
\includegraphics[width=6.35cm]{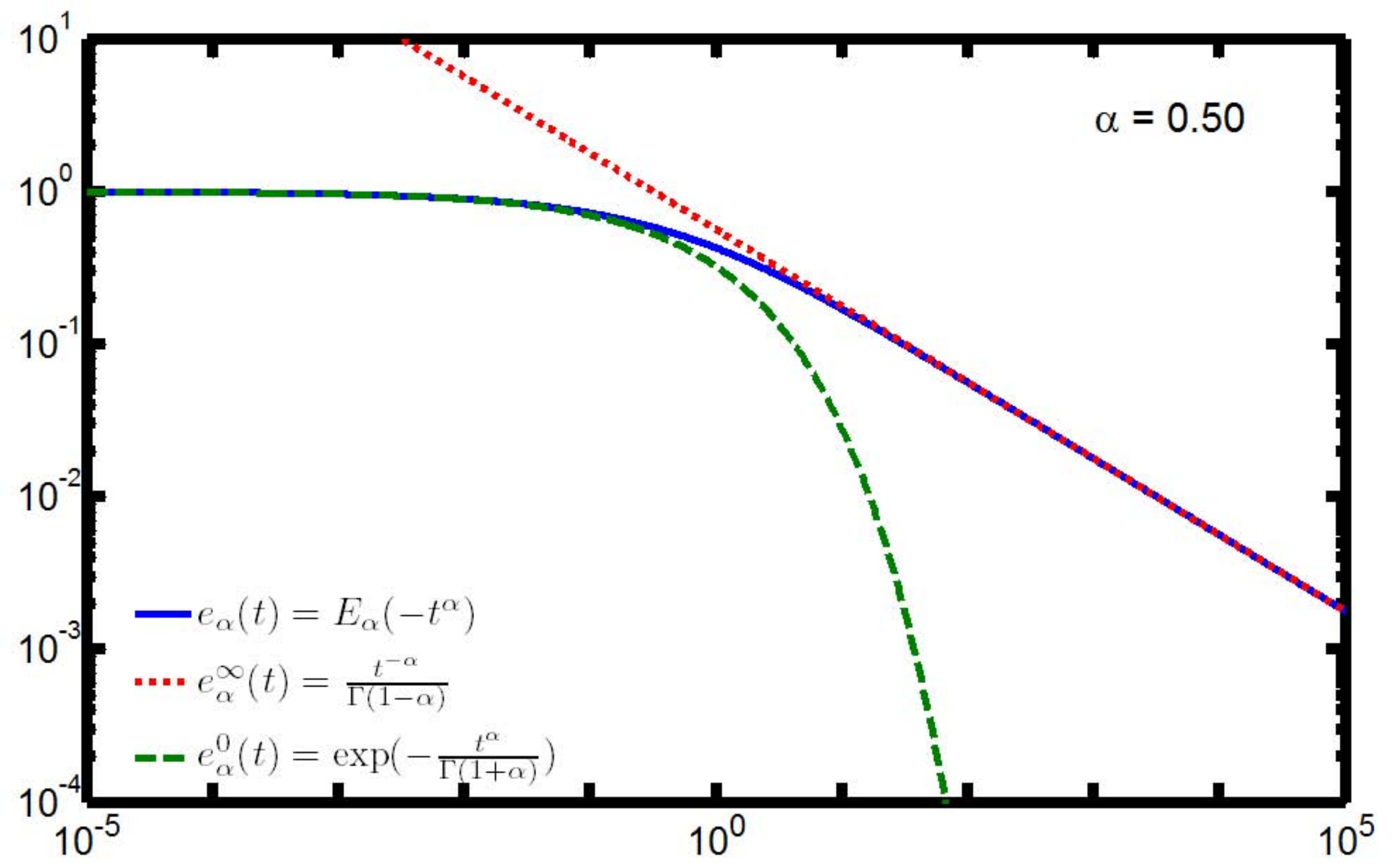}
\includegraphics[width=6.2cm]{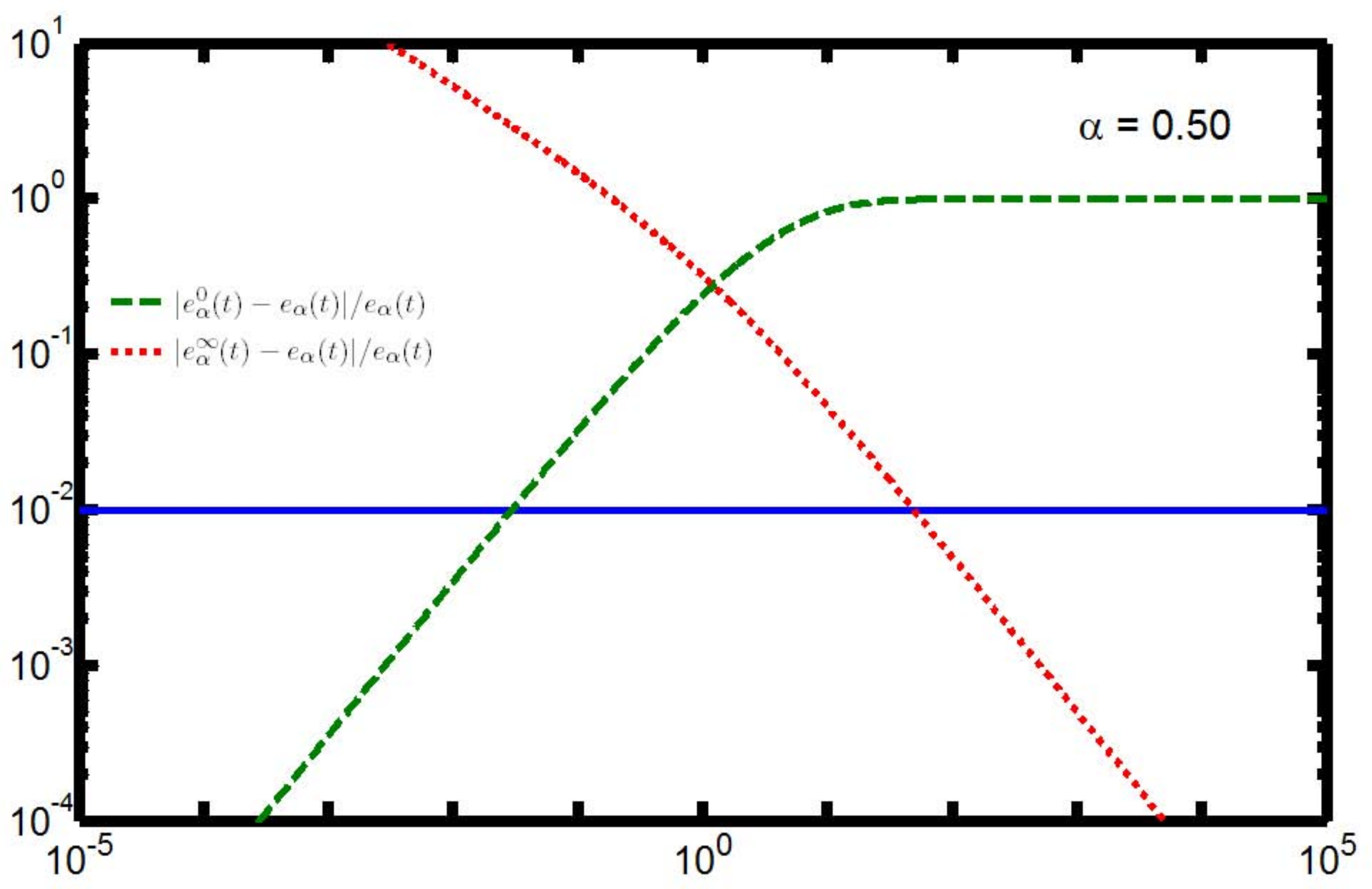}
\end{center}
{{\bf Fig.4} Approximations $e^0_\alpha(t)$ (dashed line) and $ e^\infty_\alpha(t)$ 
(dotted line) to  $e_\alpha(t)$  (LEFT) 
and  the corresponding relative errors (RIGHT)  
 in  $10^{-5} \le t \le 10^{+5}$  for $\alpha=0.50$.}
\smallskip
\begin{center}
\includegraphics[width=6.35cm]{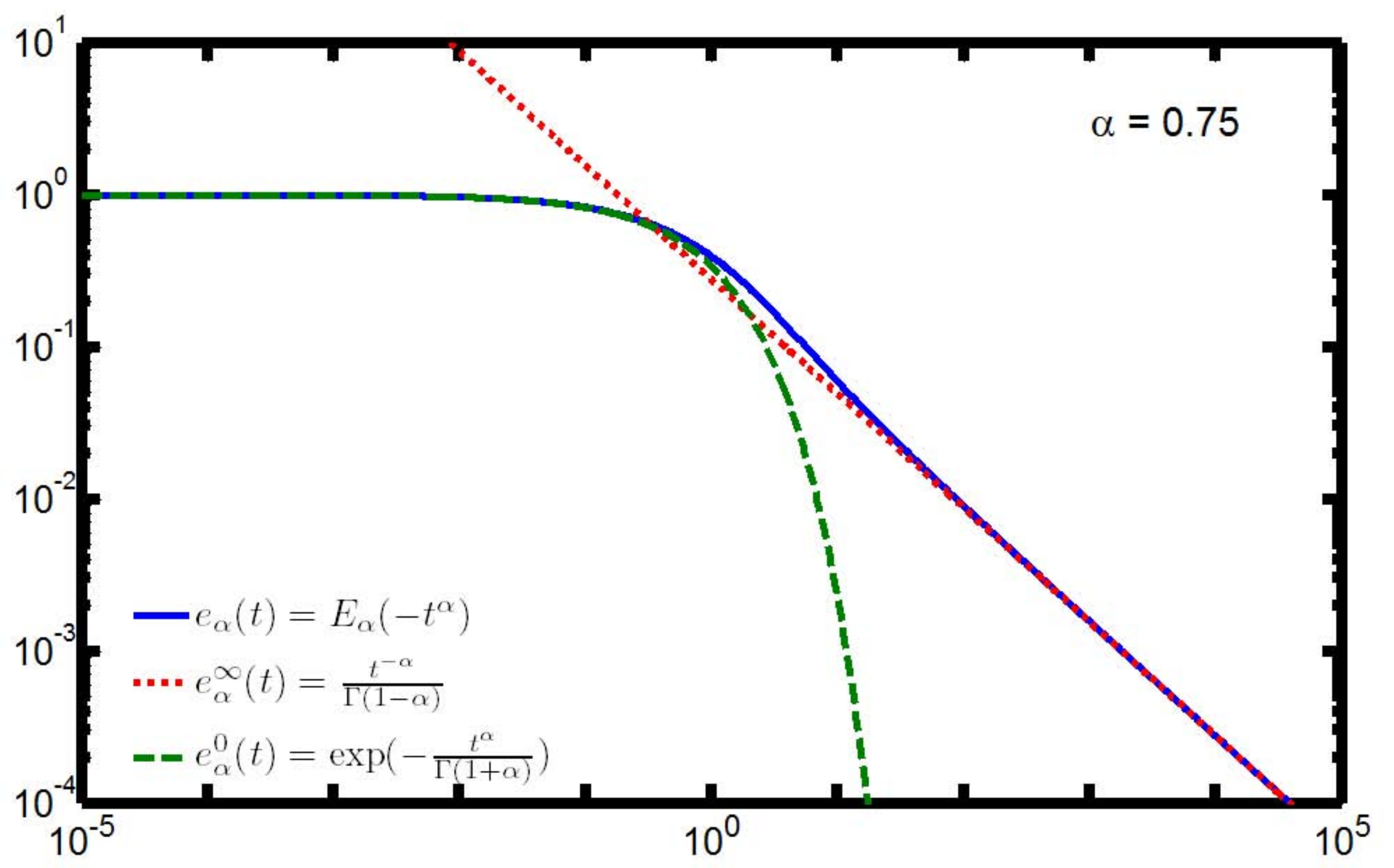}
\includegraphics[width=6.2cm]{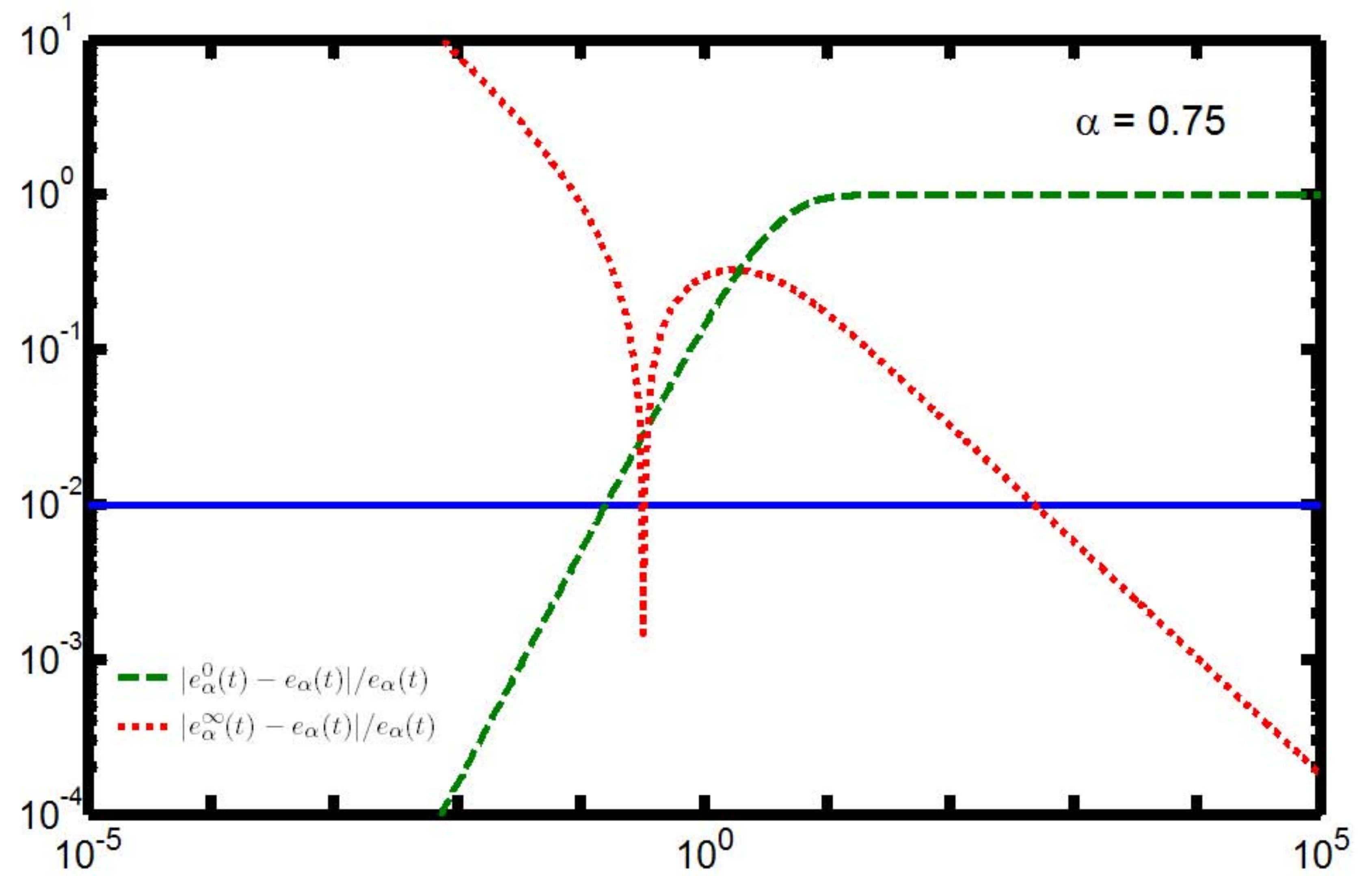}
\end{center}
{{\bf Fig.5} Approximations $e^0_\alpha(t)$ (dashed line) and $ e^\infty_\alpha(t)$ 
(dotted line) to   $e_\alpha(t)$  (LEFT) 
and  the corresponding relative errors (RIGHT)  
 in   $10^{-5} \le t \le 10^{+5}$ for $\alpha=0.75$.}
\smallskip
\begin{center}
\includegraphics[width=6.35cm]{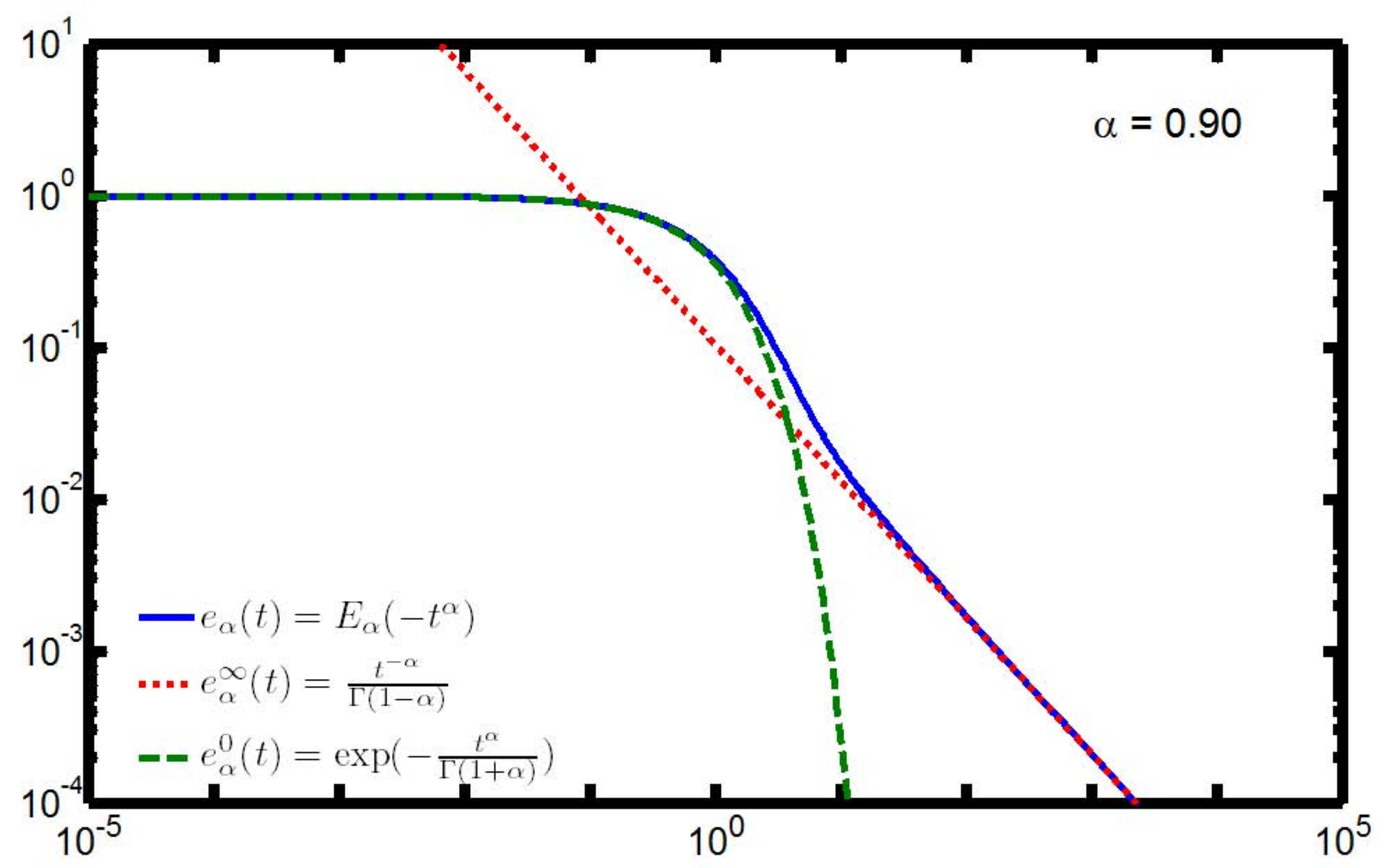}
\includegraphics[width=6.2cm]{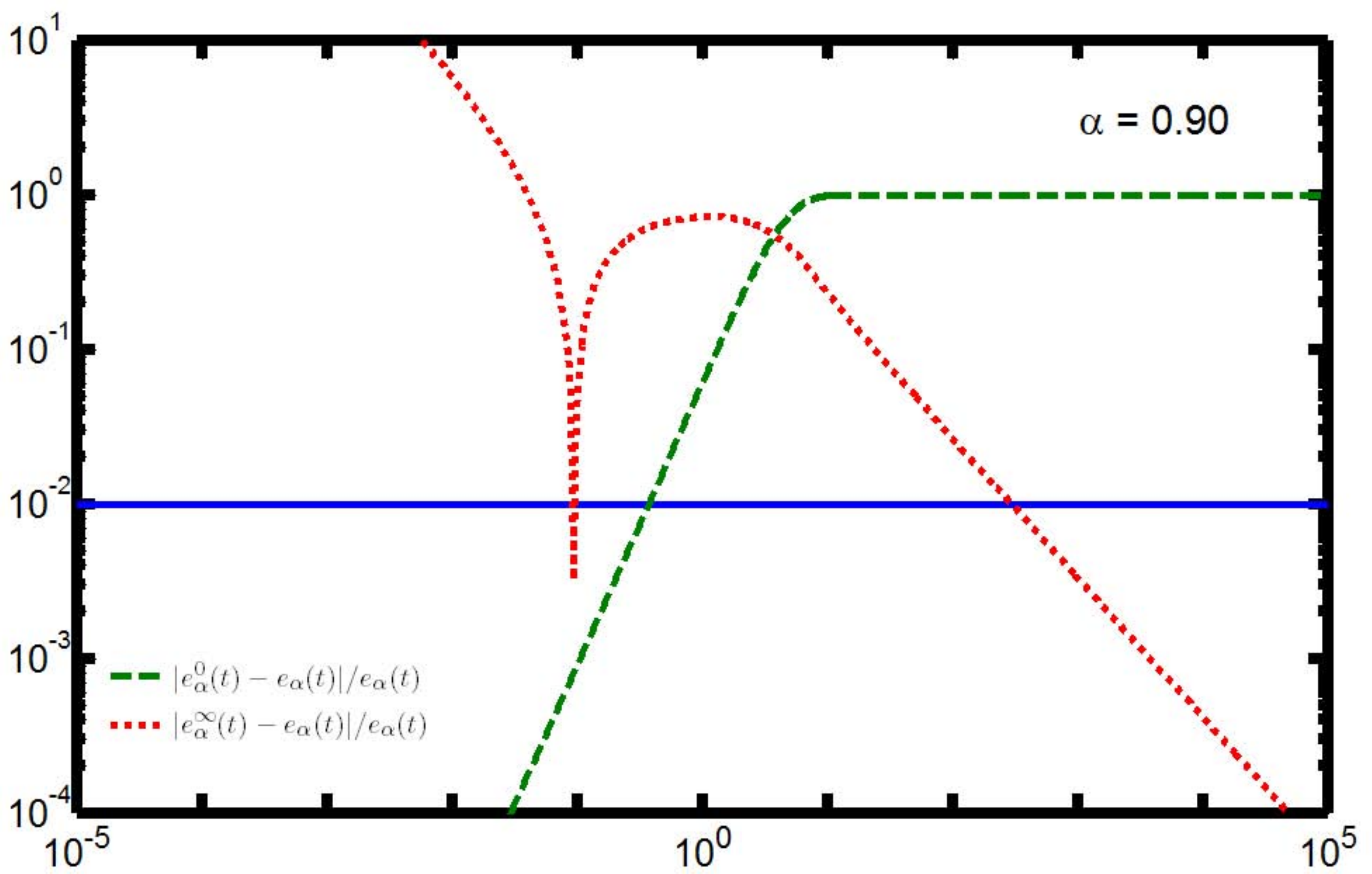}
\end{center}
{{\bf Fig.6} Approximations $e^0_\alpha(t)$ (dashed line) and $ e^\infty_\alpha(t)$ 
(dotted line) to   $e_\alpha(t)$  (LEFT) 
and  the corresponding relative errors (RIGHT)  
 in   $10^{-5} \le t \le 10^{+5}$ for $\alpha=0.90$.}

\smallskip
\begin{center}
\includegraphics[width=6.35cm]{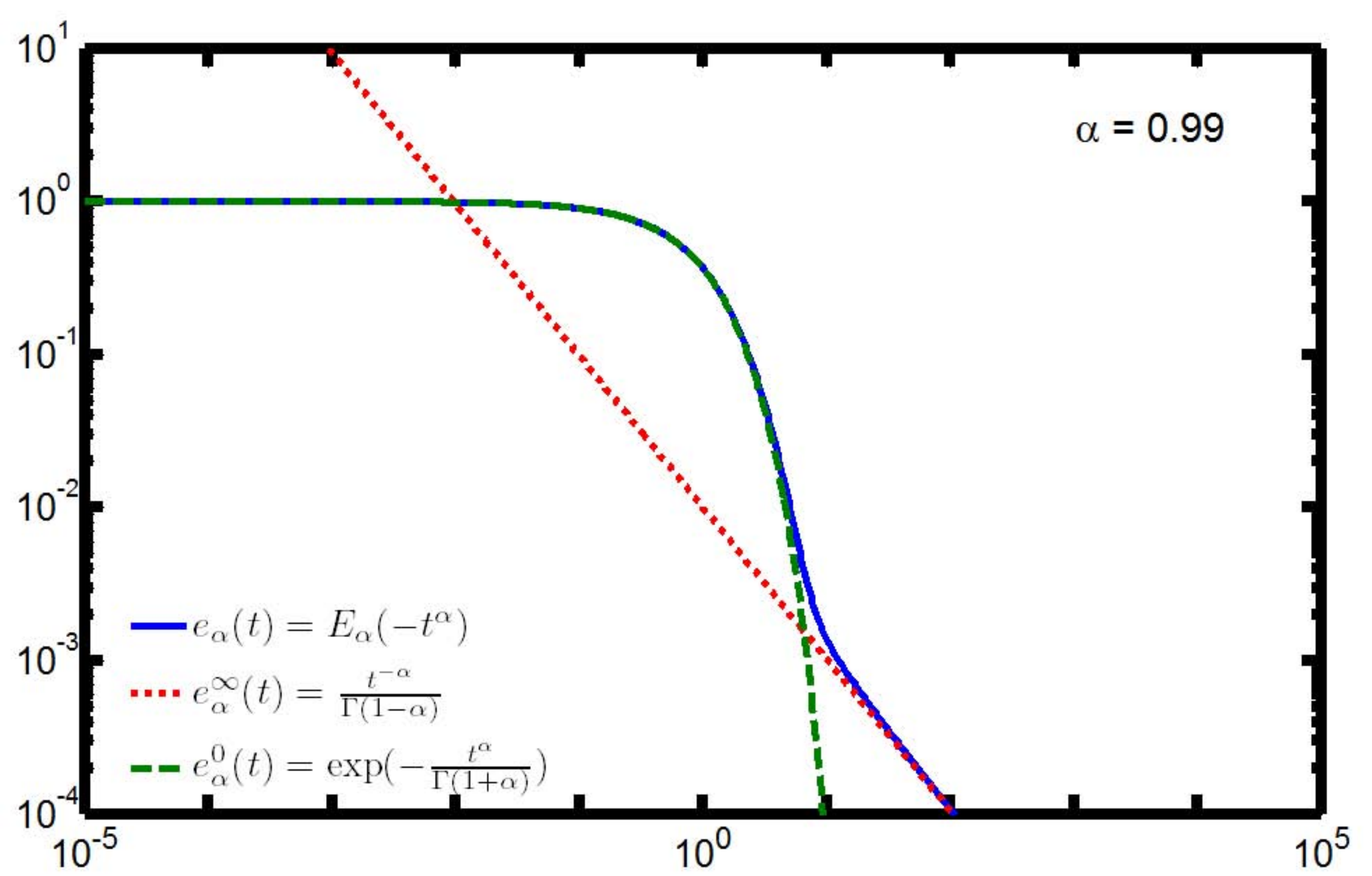}
\includegraphics[width=6.2cm]{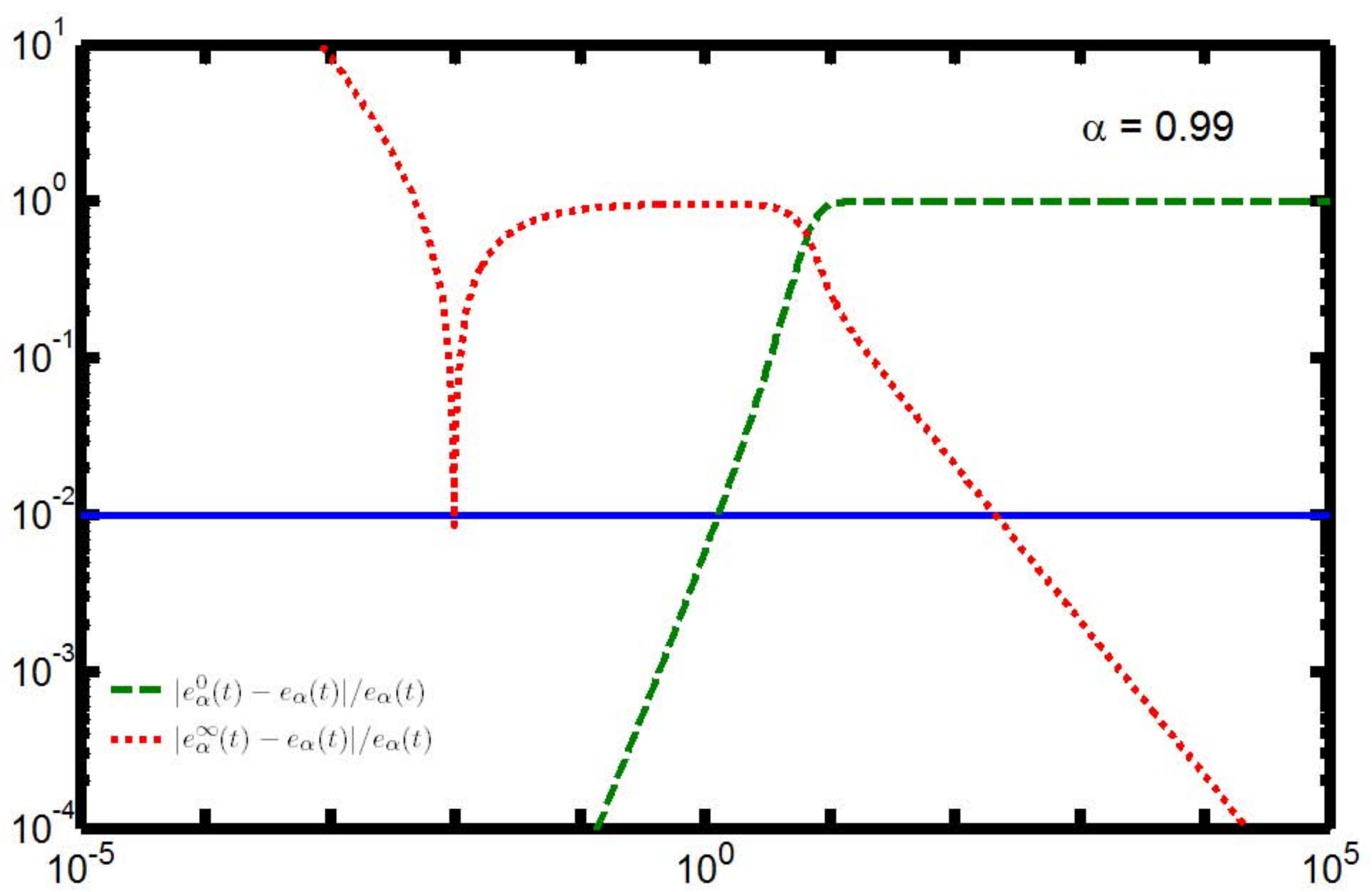}
\end{center}
{{\bf Fig.7} Approximations $e^0_\alpha(t)$ (dashed line) and $ e^\infty_\alpha(t)$ 
(dotted line) to   $e_\alpha(t)$  (LEFT) 
and  the corresponding relative errors (RIGHT)  
 in $10^{-5} \le t \le 10^{+5}$ for  $\alpha=0.99$.}
\vskip 0.15truecm
 \vsp
We note from  Figs 3-7 that, whereas the plots of $e_\alpha^0(t)$ remain always under the corresponding ones of $e_\alpha(t)$, the plots of $e_\alpha^\infty(t)$ start above those of  $e_\alpha(t)$
but, at a certain point, an intersection may occur  so changing the sign of the relative errors. 
The steep cusps occurring to the left of $t= 10^0$ in Figs 5, 6, and 7 indicates
that there the relative error is falling down to zero. 
\vsp
\subsection{The two rational asymptotic approximations} 
 We now propose  a new  set of CM functions approximating  $e_\alpha(t)$:  
 $\{ f_\alpha(t),  g_\alpha(t)\}$, alternative to  $\{ e^0_\alpha(t),  e^\infty_\alpha(t)\}$,
 obtained as the first Pad\`e approximants  $[0/1]$
  to the power series in $t^\alpha$ (2.2) and (3.2), respectively.  
For more details on the theory of Pad\`e Approximants we refer e.g. 
to   Baker \cite{Baker_BOOK1975}.
  We thus obtain the following rational functions in $t^\alpha$: 
  $$ f_\alpha(t) := {\ds \frac{1}{1+ \frac{t^\alpha}{\Gamma(1+\alpha)}}} 
  \sim 1 - \frac{t^\alpha}{\Gamma(1+\alpha)} 
  \sim e_\alpha(t)\,, \; t\to 0\,, \eqno (3.7)$$ 
   $$ g_\alpha(t) := {\ds \frac{1}{1+ t^\alpha \Gamma(1-\alpha)}}
   \sim \frac{t^{-\alpha}}{\Gamma(1-\alpha)}   
   \sim e_\alpha(t)\,,  \; t\to \infty\,.\eqno (3.8)$$ 
   Now we prove the inequality 
   $$ g_\alpha(t) \le f_\alpha(t)\,, \quad t\ge 0 \,, \quad 0<\alpha<1\,, \eqno (3.9)$$
   as a straightforward consequence of the reflection formula of the gamma function.   
   In fact, recalling the definitions  (3.7)-(3.8) we have for $\forall t \ge 0$ and $\alpha \in (0,1)$:
   $$   g_\alpha(t) \le f_\alpha(t)\, \Longleftrightarrow\, \Gamma (1-\alpha) \ge 
   \frac{1}{\Gamma(1+\alpha)}\, \Longleftrightarrow \, \Gamma(1-\alpha) \, \Gamma(1+\alpha)
   = \frac{\pi \alpha}{\sin (\pi \alpha)} \ge 1\,.  
    $$
In  Figs  8-12 LEFT  we compare for $\alpha = 0.25, 0.5, 0.75, 0.90, 0.99$ in   logarithmic scales  the function $e_\alpha(t)$ (continuous line) and its rational asymptotic representations,  $f_\alpha(t)$ valid for $ t\to 0$ (dashed line)  and $g_\alpha(t)$  valid for $t\to \infty$ (dotted line).
We have chosen the time range  $10^{-5} \le t \le 10^{+5}$.
In the RIGHT we have shown  the plots of the relative errors (no longer in absolute values)
$$  \frac{f_\alpha(t)- e_\alpha(t)}{e_\alpha(t)}\,, \quad
\frac{e_\alpha(t)- g_\alpha(t|}{e_\alpha(t)}\,, \eqno (3.10)$$
pointing out a continuous line at an error 1\% under which the approximation can be considered reliable.    
\vsp
\begin{center}
\hskip -0.5truecm
\includegraphics[width=6cm]{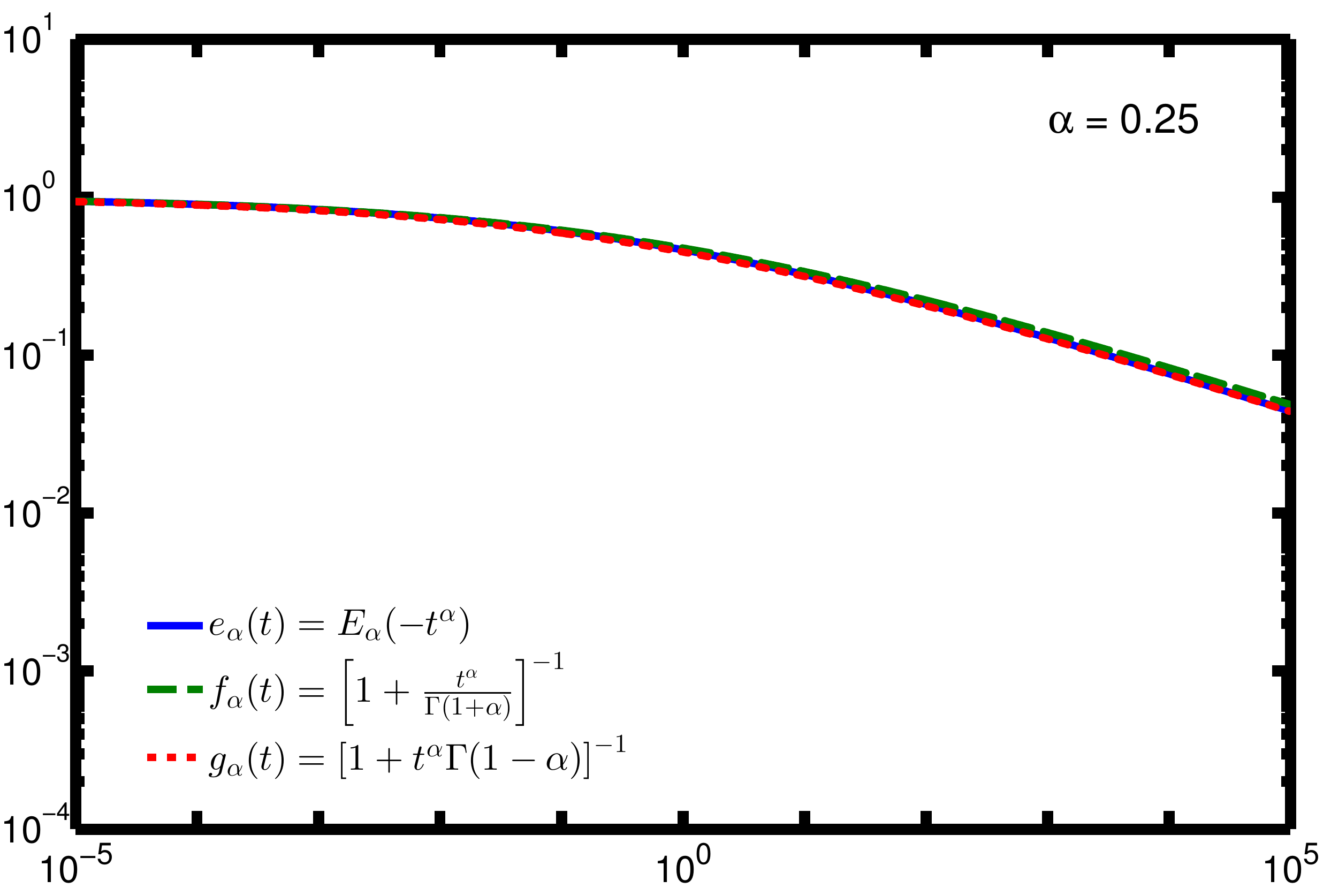} 
\includegraphics[width=6cm]{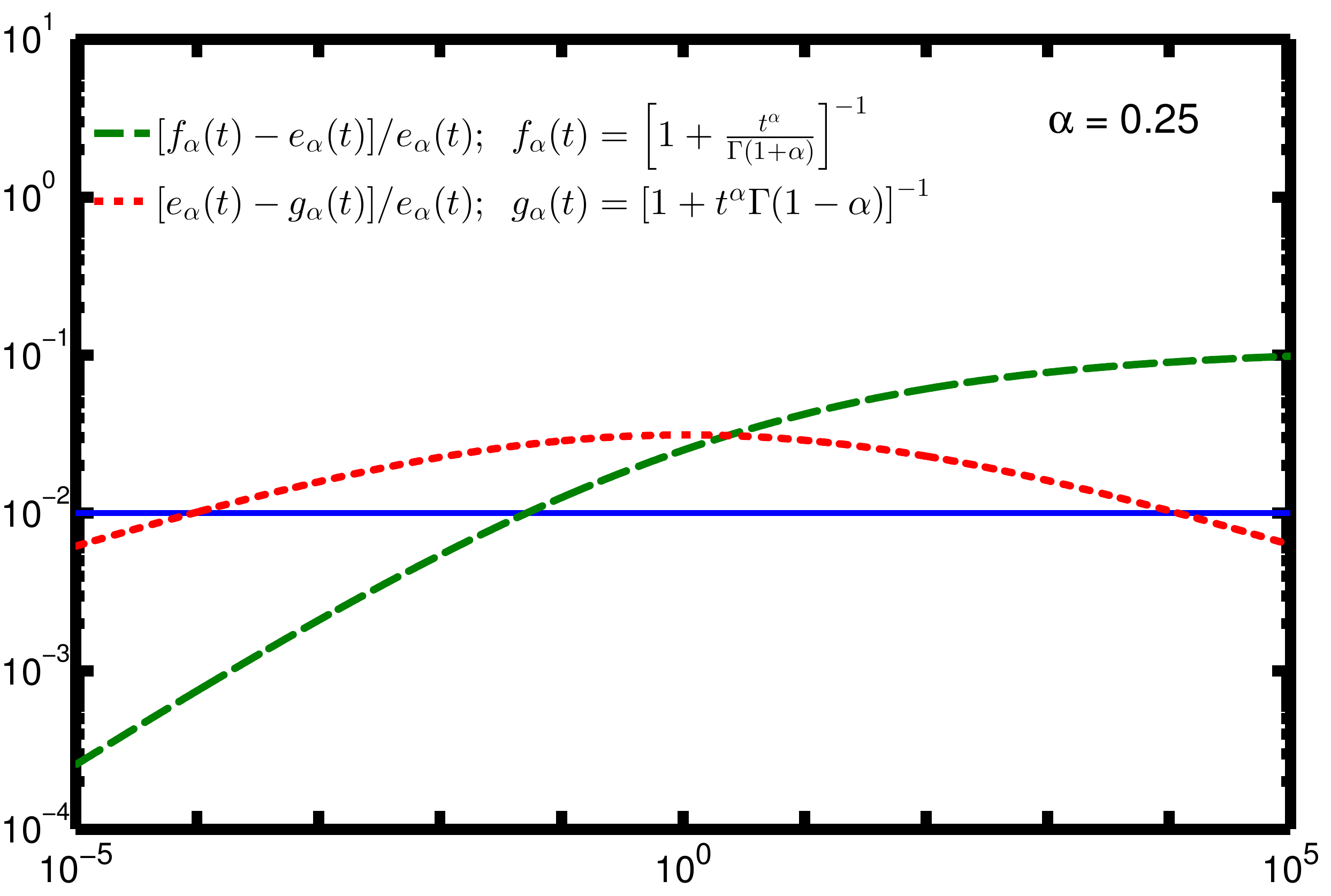}
\end{center}
{{\bf Fig.8} Approximations $f_\alpha(t)$ (dashed line)  and $ g_\alpha(t)$ (dotted line) 
to   $e_\alpha(t)$  (LEFT) 
and  the corresponding relative errors (RIGHT)  
 in  $10^{-5} \le t \le 10^{+5}$  for $\alpha=0.25$.}

\begin{center}
\hskip -0.5truecm
\includegraphics[width=6cm]{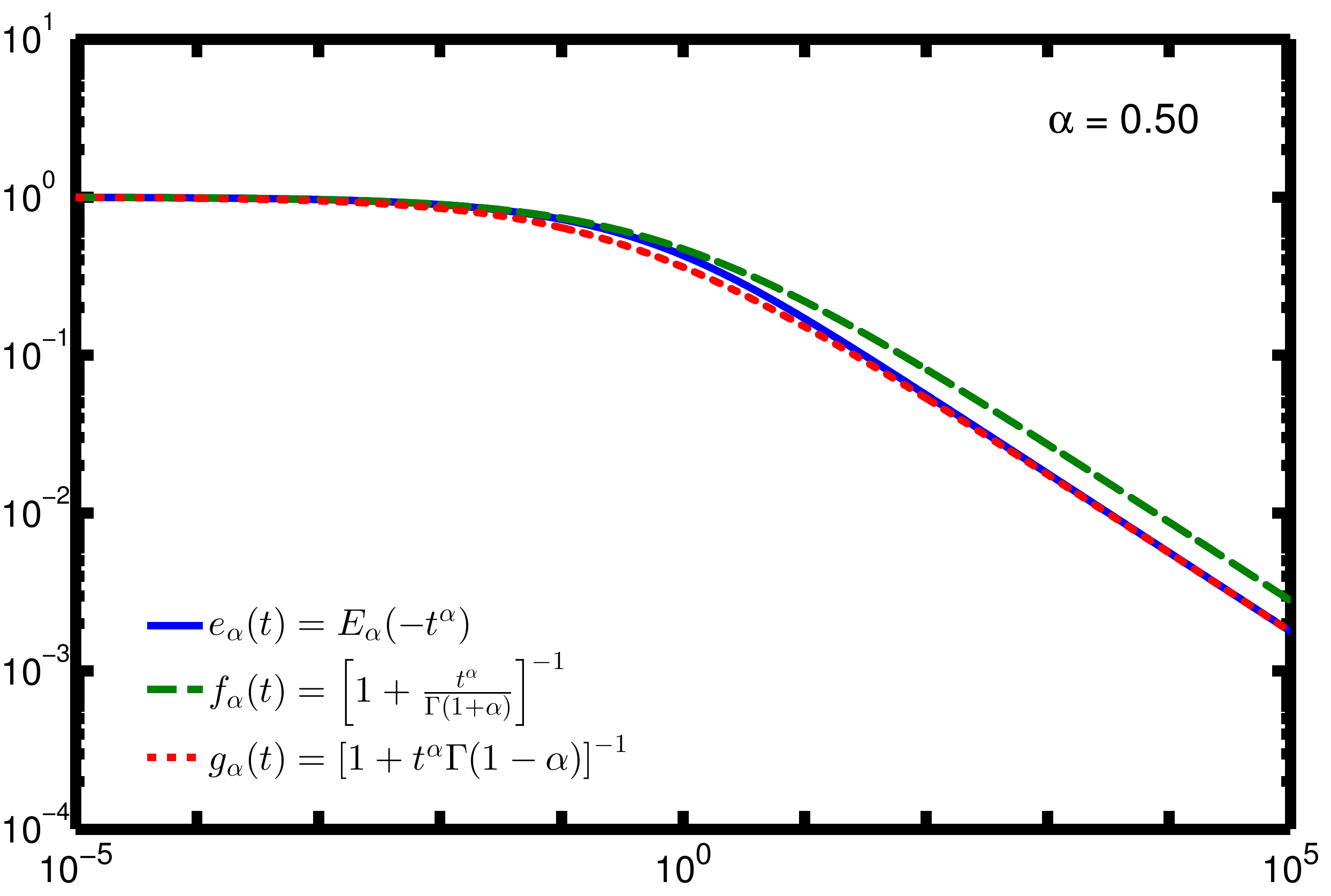}
\includegraphics[width=6cm]{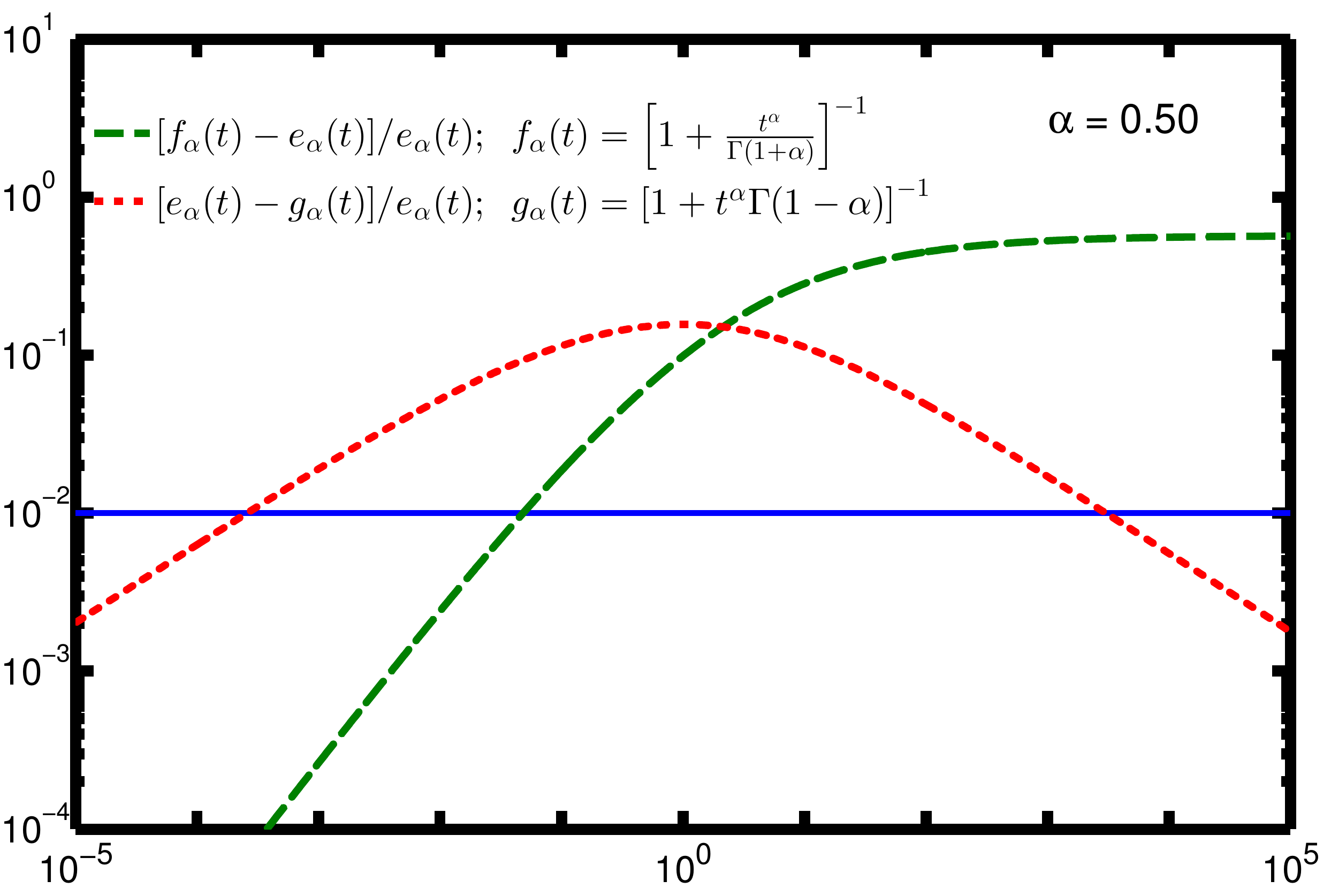}
\end{center}
{{\bf Fig.9} Approximations $f_\alpha(t)$ (dashed line) and $ g_\alpha(t)$ (dotted line) 
to   $e_\alpha(t)$  (LEFT) 
and  the corresponding relative errors (RIGHT)  
 in     $10^{-5} \le t \le 10^{+5}$ for $\alpha=0.50$.}
\newpage
\begin{center}
\hskip -0.5truecm
\includegraphics[width=6cm]{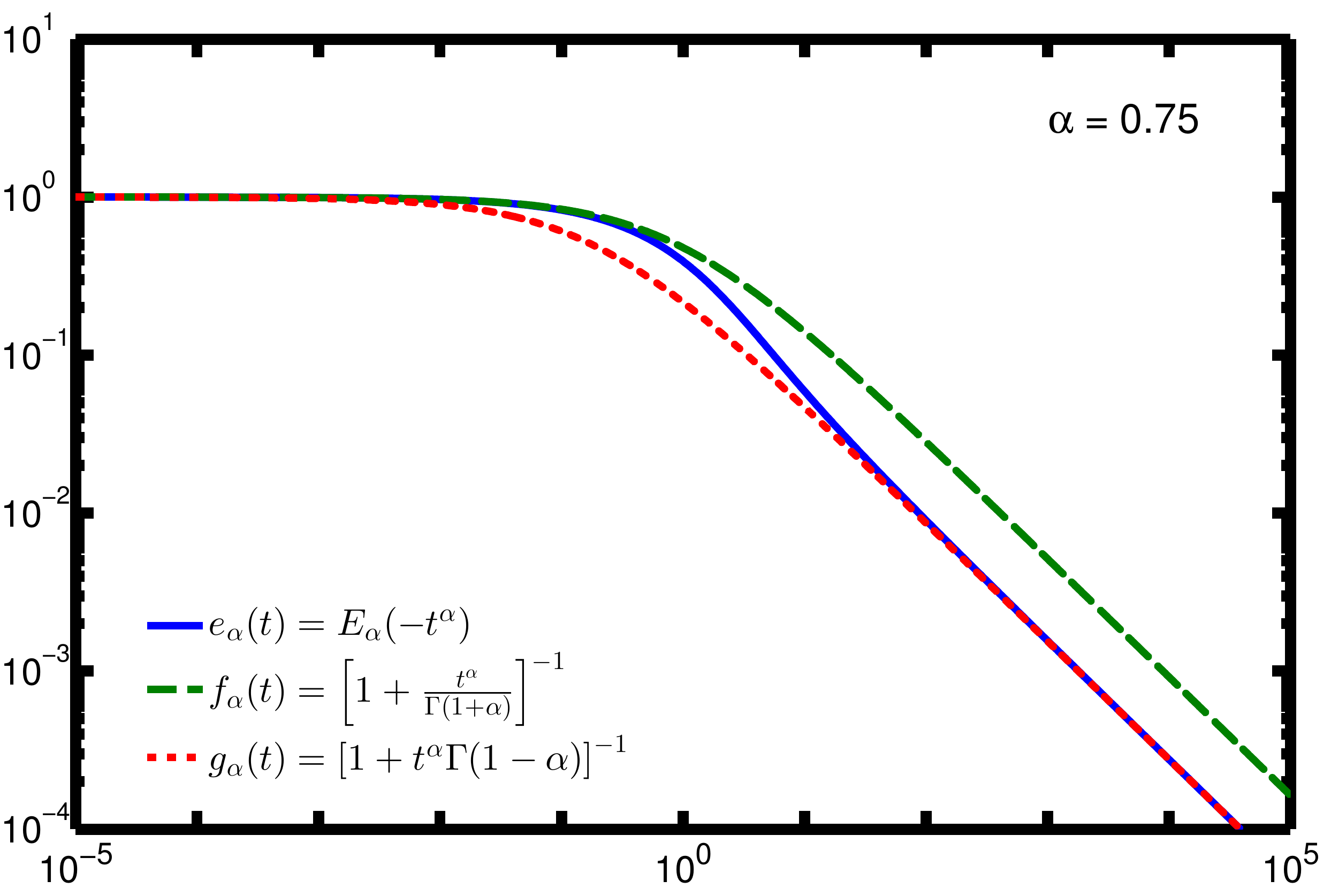}
\includegraphics[width=6cm]{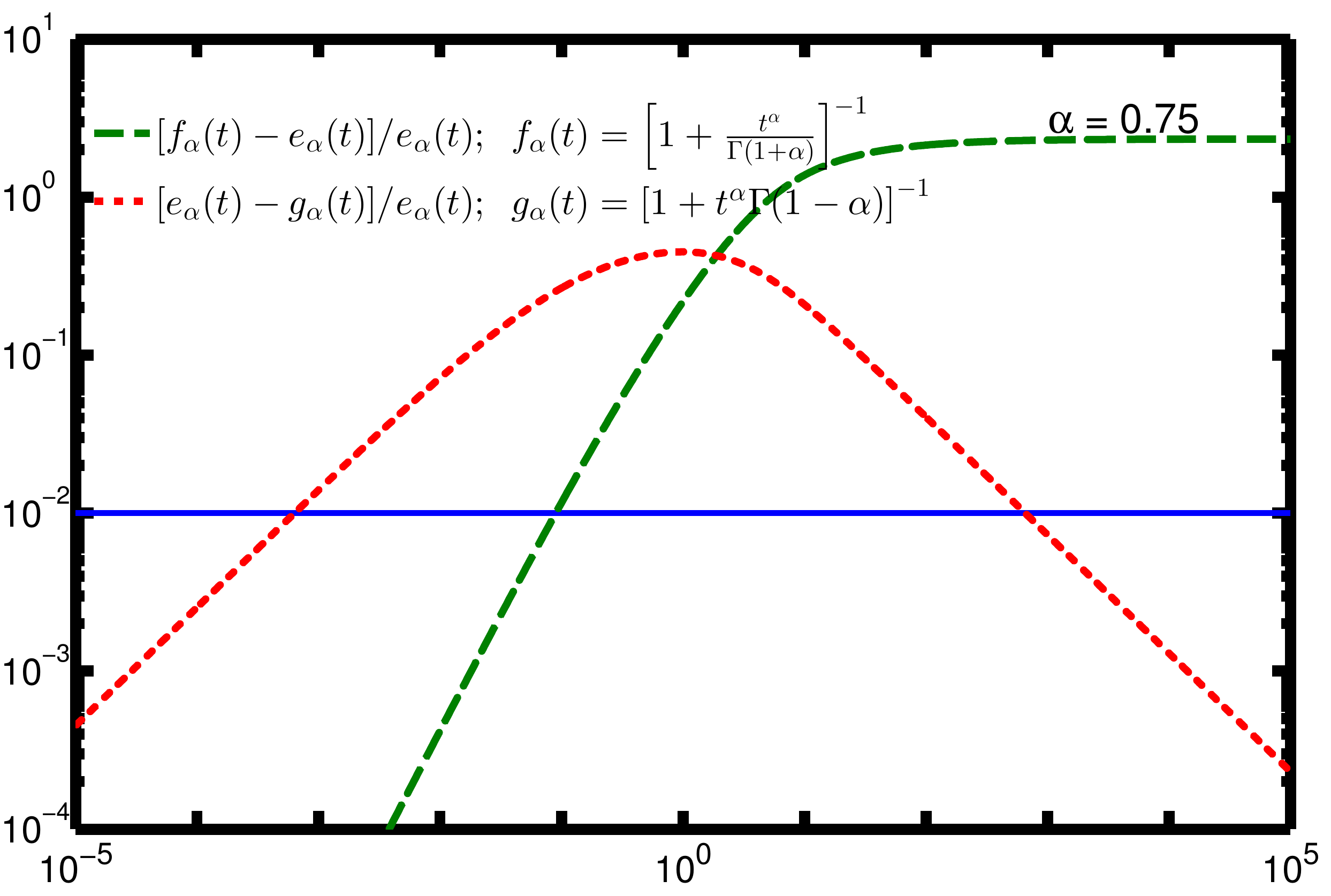} 
\end{center}
{{\bf Fig.10} Approximations $f_\alpha(t)$ (dashed line) and $ g_\alpha(t)$ (dotted line) 
to  $e_\alpha(t)$  (LEFT) 
and  the corresponding relative errors (RIGHT)  
 in   $10^{-5} \le t \le 10^{+5}$  for $\alpha=0.75$.}
\smallskip
\begin{center}
\hskip -0.5truecm
\includegraphics[width=6cm]{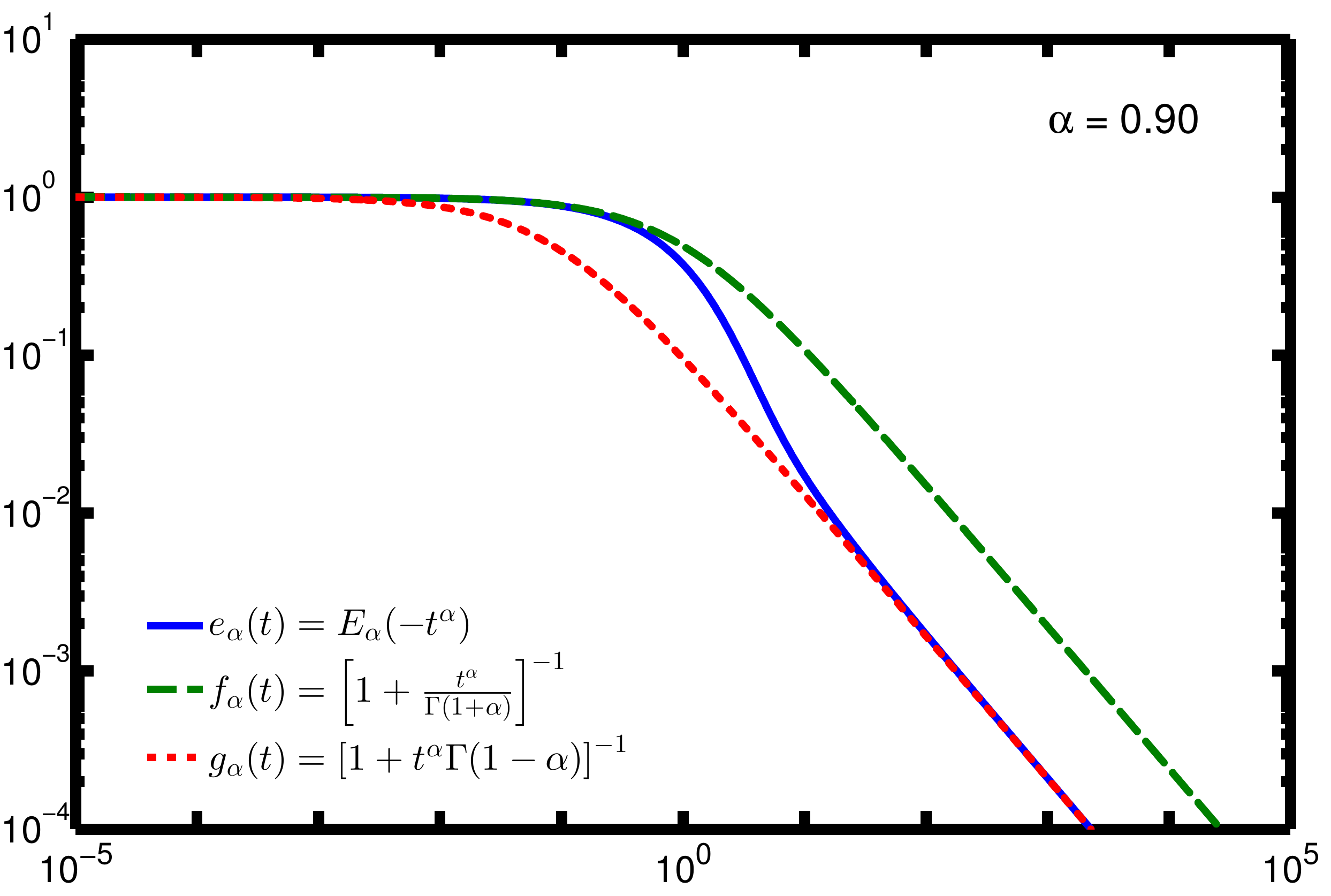} 
\includegraphics[width=6cm]{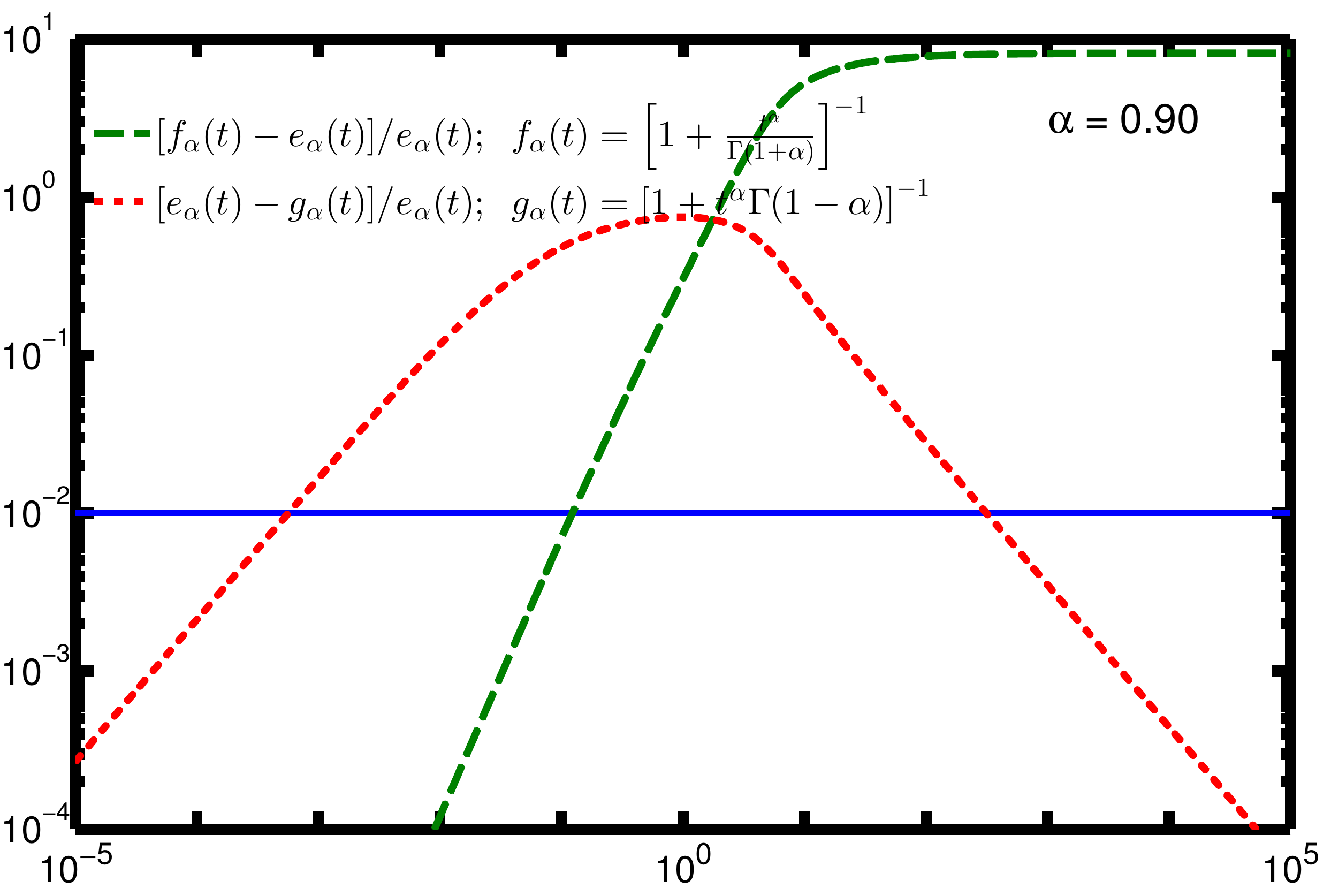}
\end{center}
{{\bf Fig.11} Approximations $f_\alpha(t)$ (dashed line) and $ g_\alpha(t)$ (dotted line) 
to  $e_\alpha(t)$  (LEFT) 
and  the corresponding relative errors (RIGHT)  
 in   $10^{-5} \le t \le 10^{+5}$  for $\alpha=0.90$.}
\smallskip
\begin{center}
\hskip -0.5truecm
\includegraphics[width=6cm]{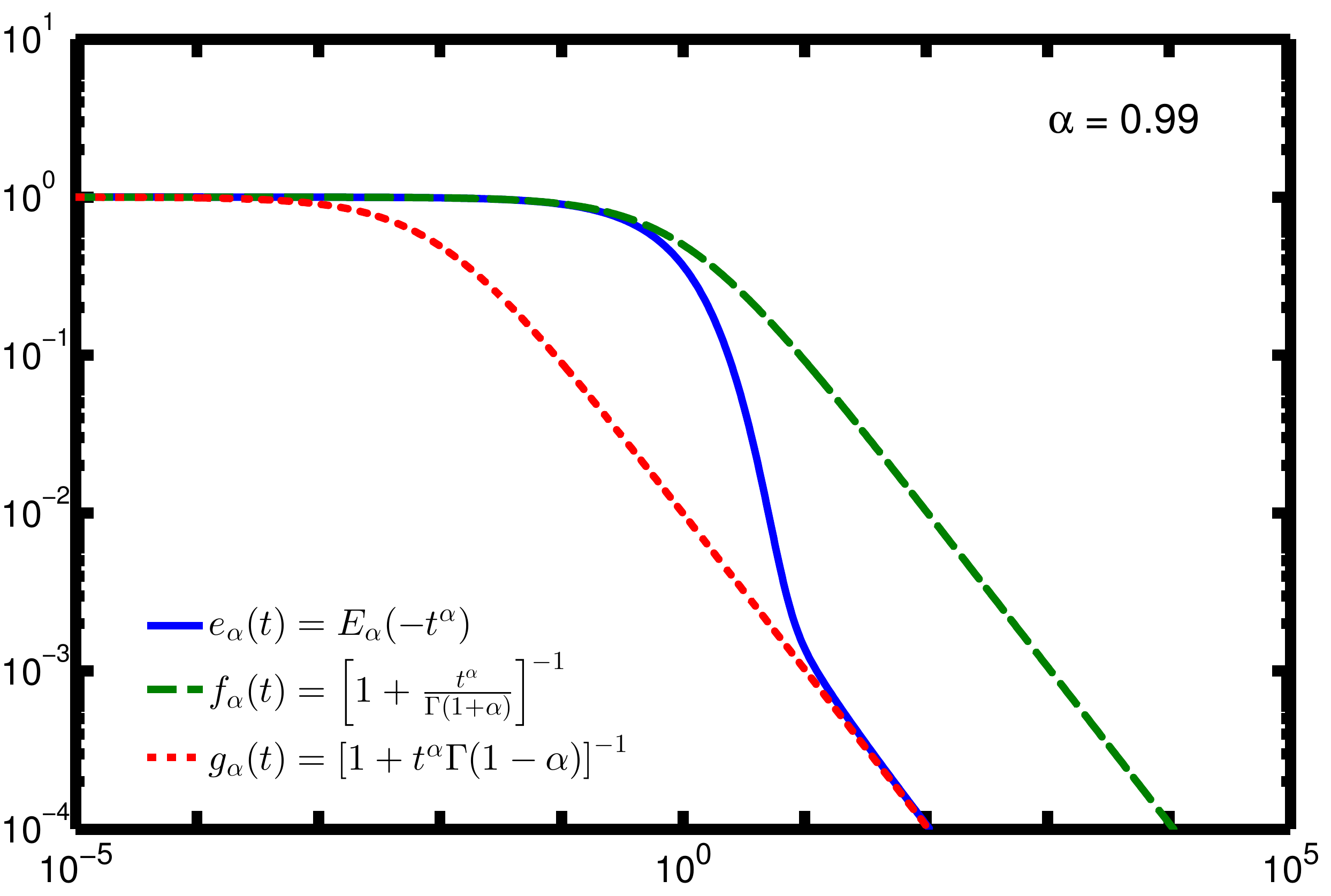} 
\includegraphics[width=6cm]{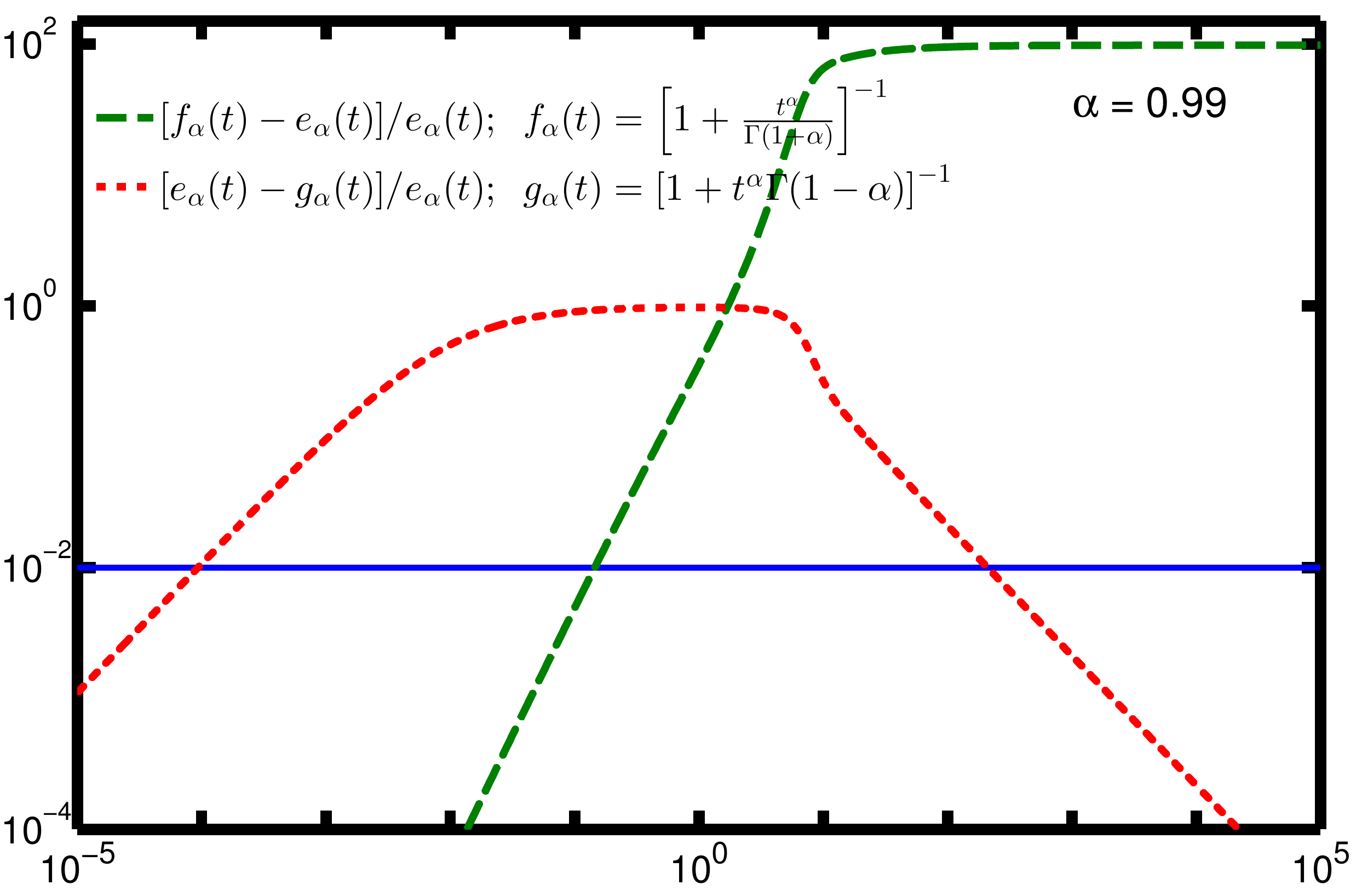}
\end{center}
{{\bf Fig.12} Approximations $f_\alpha(t)$ (dashed line) and $ g_\alpha(t)$ (dotted line) 
to   $e_\alpha(t)$  (LEFT) 
and  the corresponding relative errors (RIGHT)  
 in   $10^{-5} \le t \le 10^{+5}$  for $\alpha=0.99$.}

\vskip 0.15truecm
\vsp 
As a matter of fact, from the plots in Figs 8-12,  
 we recognize, for the  time range and for  $\alpha\in (0,1)$ considered by us,   
  the relevant inequality 
$$ g_\alpha(t) \le e_\alpha(t) \le f_\alpha(t) \,, \eqno(3.11)$$
that is  $g_\alpha(t)$ and $f_\alpha(t)$ provide lower and upper bounds 
to our Mittag-Leffler function $e_\alpha(t)$ . 
 This of course can be seen as a conjecture that we leave as an open problem to be proved 
 (or disproved) by specialists of CM functions.
 \vsp
 We also see  that whereas the     short  time approximation $f_\alpha(t)$ turns out to be good  only for small times,  the long time approximation $g_\alpha(t)$ is good (surprisingly) also for short times,
falling down only in an intermediate time range. 
In fact from the RIGHT of  Figs 8-12  we can estimate  the ranges of validity
when the relative error is less than $1\%$.  We could further  discuss on them.
\vsp 
Concerning Pad\`e  Approximants  (PA) for the Mitag-Leffler functions
$E_\alpha(-x)$ for $x>0$ and $0<\alpha<1$  
 we like  to cite  Freed et al. \cite{Freed_NASA2002}for  an appendix
 devoted to the table of PA, 
 Starovoıtov and  Starovoıtova
\cite{Staroitov-Staroitova_SB2007}
for theorems on  the best uniform rational approximations.
Quite recently Zeng \& Chen \cite{Zeng-Chen_PRE-PRINT2013}
have  found  a global Pad\`e approximation with degree 2 of the generalized
Mittag-Leffler  function $E_{\alpha, \beta}(−x)$ 
with  $x>0$, $0<\alpha<1$ and $\beta \ge \alpha$.
In this interesting pre-print   the uniform approximation can account
for both the Taylor series for small arguments and asymptotic series for large arguments, but no upper and lower bounds are obtained as in our case. 

We point out the fact  that our first PA's  [0/1] are constructed from two different  series (in positive and negative powers of $t^\alpha$)  both resulting   CM functions  whereas  the successive PA's of higher order  can exhibit oscillations for $t>0$.  

\section{Conclusions}
We have discussed some noteworthy properties of the Mittag-Leffler function $E_\alpha(-t^\alpha)$
with $0<\alpha<1$  in the range $t>0$.
\vsp
Being a completely monotone (CM) function, because of the Bermstein theorem this function admits  non-negative  frequency  and time spectra. We have pointed out that these two spectra  are equal, so providing a universal scaling property.
\vsp
Furthermore, in view of a numerical approximation  we have compared two different sets of approximating CM functions, asymptotically  equivalent to   $E_\alpha(-t^\alpha)$ for small  and large times: the former commonly used in the literature, the latter probably new. The last set is constituted by two simple rational   
functions that provide upper and lower bounds, at least in our numerical examples on a large time range.
\vsp
We are allowed to the conjecture that this bounding property is always valid    for any $t>0$ and 
for any $\alpha \in (0,1)$: an open problem offered to  specialists of CM functions.
\vsp
Last but not the least, we note that our  rational function approximating $E_\alpha(-t^\alpha)$ for large times  provides   (surprisingly) a good numerical approximation also for short times, so failing only in an intermediate range.
\vsp

\section*{Final remark}
Since the appearance of the first version (May 2013), this paper was submitted to the attention of some colleagues  and to any interested reader of arXiv  in order to get the proof of our conjecture. 
Only recently (October 2013) a proof  was provided by  Thomas Simon
(University of Lille, France)  based on probability arguments, see \cite{Simon_2013}.
We note, however, that in July 2013
Renato Spigler (University of Rome 3, Italy)  proved, in part, the conjecture
made in eq. (3.11), for $x := t^{\alpha}$ in some right neighborhood of
$x=0$. This set is the most critical in establishing such conjecture,
see Fig.s 8 -- 12.
See R. Spigler, 2013 SIAM Annual Meeting, July 8-12, 2013, San Diego, CA
(USA), MS 119, Mini-Symposium on ``Special Functions: Applications and
Numerical Aspects'' -- Part II of II, p. 100.


\subsection*{Acknowledgments} 
The   author is grateful to his students, Marco Di Cristina and Andrea Giusti,  for their valuable help in   plotting the Mittag-Leffler function.  They have used the MATLAB routine by Podlubny  \cite{Podlubny_MATLAB2006}
that is essentially based on the MATHEMATICA routine   
by Gorenflo et al. \cite{Go-Lou-Lu_FCAA2002}.
Furthermore the author has appreciated constructive remarks and suggestions of 
Rudolf Gorenflo (Emeritus Professor of Mathematics, Free University Berlin, Germany)  that helped to improve the first version of this paper. 




\input{MAINARDI_AIMS-BIBLIO.tex}

\end{document}

%% file: MAINARDI_AIMS-BIBLIO.tex

%% file: MAINARDI_M-L-4REV_ARXIV-FABRIZIO.bbl
\begin{thebibliography}{99}

\bibitem{Baker_BOOK1975}
G.A. Baker,  
``{Essentials of  Pad\`e Approximants}'',
Academic Press, New York, 1975.

\bibitem{BDST_BOOK2012}
D. Baleanu,  K.  Diethelm, E.  Scalas  and  J. J.  Trujillo,  
``{Fractional Calculus. Models and Numerical Methods}'',
World Scientific, Singapore, 2012.


\bibitem{Beghin-Orsingher_EPJ2010} 
L. Beghin  and E. Orsingher,  
\emph{Poisson-type processes governed by fractional and
higher-order recursive differential equations},
{Electronic J. Probability}, {\bf 15} (2010),  Paper no. 22,  684-–709.

\bibitem{Capelas-et-al_EPJ-ST2011}
E. Capelas de Oliveira, F. Mainardi  and J. Vaz Jr,  
{\it Models based on Mittag-Leffler functions for anomalous relaxation in dielectrics},
{Eur. Phys. J., Special Topics}, {\bf 193} (2011) 161--171.
[E-print {\tt arxiv.org/abs/1106.1761}]


\bibitem{Caputo-Mainardi_PAGEOPH1971}
M. Caputo  and  F.  Mainardi,   
   {\it A new dissipation model based on memory mechanism},
  {Pure and Appl. Geophys. (PAGEOPH)}, {\bf 91} (1971), 134--147.
  [Reprinted in {Fract. Calc.  Appl. Anal.},{\bf 10} No 3 (2007), 309--324]

\bibitem{Caputo-Mainardi_RNC1971}
 M. Caputo  and  F. Mainardi,  
  {\it Linear models of dissipation in  anelastic solids},
  {Riv. Nuovo Cimento} (Ser. II), {\bf 1} (1971), 161--198.




\bibitem{Cole-Cole_1942}
 K. S. Cole  and  R. H. Cole, 
 {\it Dispersion and absorption in dielectrics, II. Direct current characteristics},
{J. Chemical Physics}, {\bf 10} (1942), 98--105.

\bibitem{Davis_BOOK1936}
 H.T. Davis,  
``{The Theory of Linear Operators}'',
  The Principia Press, Bloomington, Indiana, 1936.

\bibitem{Diethelm_BOOK2010}
K. Diethelm,  
``{The Analysis of Fractional Differential Equations}'',
Springer, Lecture Notes in Mathematics No 2004,  Heidelberg, 2010. 


\bibitem{Dzherbashyan_BOOK1966}
 M. M. Dzherbashyan, , 
 ``{Integral Transforms and  Representations of Functions in the Complex Plane}'',
   Nauka, Moscow., 1966  [in Russian]. 

\bibitem{Erdelyi_HTF1955}
 A. Erd\'elyi, W. Magnus,. F. Oberhettinger and F.G. Tricomi, 
``{Higher Transcendental Functions}'', Vol.3 ,
Ch. 18: Miscellaneous Functions: pp. 206-227.
 McGraw-Hill, New York, 1955. 


\bibitem{Feller_BOOK1971}
W.  Feller, 
``{An Introduction to Probability Theory and its Applications}'',
Vol. II, Second Edition,
  Wiley, New York, 1971. 



\bibitem{Freed_NASA2002}
 A. Freed, K. Diethelm  and Yu.  Luchko,  
 {\it Fractional-order Viscoelasticity (FOV): Constitutive Development using
 the Fractional Calculus}, 
  {First Annual Report, NASA/TM-2002-211914},
  Gleen Research Center, 2002, pp. XIV -- 121.
  


\bibitem{Go-Lou-Lu_FCAA2002}
R. Gorenflo, J. Loutchko  and Yu.  Luchko,  
\emph{Computation of the Mittag-Leffler function and its derivatives},
{Fract. Calc. Appl. Anal.}, {\bf 5} (2002), 491--518.

\bibitem{Gorenflo-Mainardi_CISM1997}
R.  Gorenflo  and F. Mainardi,  
  \emph{Fractional calculus: integral and differential equations of fractional order},
   in   ``{Fractals and Fractional Calculus in Continuum Mechanics}''
   (eds. A. Carpinteri and  F. Mainardi),
 Springer Verlag, Wien, 1997, pp. 223--276.
 [E-print {\tt arxiv.org/abs/0805.3823}]

\bibitem{Gross_JAP1947}
 B. Gross,  
\emph{On creep and relaxation},
{J. Appl. Phys.}, {\bf 18} (1947), 212--221.

 \bibitem{Hilfer_BOOK2000}
 R. Hilfer (editor),  
``{Fractional Calculus, Applications in Physics}'',
World Scientific, Singapore, 2000.


\bibitem{Hille-Tamarkin_1930}
E. Hille  and J. D. Tamarkin,  
  \emph{On the theory of linear integral equations},
  {Ann. Math.},  {\bf 31} (1930), 479--528.

 \bibitem{Kilbas-et-al_FCAA2013}
A. A. Kilbas,  A . A. Koroleva  and S. V. Rogosin,  
\emph{Multi-parametric Mittag-Leffler functions and their extensions},
{Fract. Calc. Appl. Anal.}, {\bf  16} No 2 (2013), 378--404.
(DOI: 10.2478/s13540-013-0024-9)


\bibitem{Kilbas-Saigo_1995}
A. A. Kilbas and M.  Saigo,
{\it  On solution of integral equations of Abel-Volterra type}, 
{Differential and Integral Equations},
\textbf{8}, No. 5 (1995), 993--1011.

\bibitem{Kilbas-Saigo_BOOK2004}          
 A. A. Kilbas and M.  Saigo,  
``{$H$-Transforms. Theory and Applications}'', 
Chapman and Hall/CRC, Boca Raton, FL, 2004.


 \bibitem{Kilbas-Srivastava-Trujillo_BOOK2006}
A. A. Kilbas, H. M  Srivastava and   J. J. Trujillo,  
``{Theory and Applications of Fractional Differential Equations}'',
Elsevier, Amsterdam, 2006. 


\bibitem{Kiryakova_BOOK1994}
V. Kiryakova,  
``{Generalized Fractional Calculus and Applications}'',
Longman \& J. Wiley, Harlow -  New York, 1994.


\bibitem{Kiryakova_CMA2010}
V. Kiryakova, 
\emph{The multi-index Mittag-Leffler functions as important
class of special functions of fractional calculus},
 {Comp. Math. Appl.}, {\bf  59} No 5,  (2010), 1885--1895. 

\bibitem{Kiryakova-Luchko_2010}
V.  Kiryakova  and Yu.  Luchko,  
 \emph{The multi-index Mittag-Leffler functions and their applications for solving fractional order problems in applied analysis},
In: American Institute of Physics - Conf. Proc. Vol. 1301, 
Proc. AMiTaNS'10 (2010), 597--613 (doi:10.1063/1.3526661)

\bibitem{KLM_BOOK2012}
J. Klafter,  S. C.  Lim,   and R. Metzler  (Editors), 
``{Fractional Dynamics,  Recent Advances}'', 
World Scientific, Singapore, 2012.


\bibitem{Magin_BOOK2006}
R .L Magin,  
``{Fractional Calculus in Bioengineering}'',
Begell House Publishers, Connecticut, 2006.


\bibitem{Mainardi_BOOK2010} 
F. Mainardi, 
 ``{Fractional  Calculus and Waves in Linear Viscoelasticity}''
  Imperial College Press, London and World Scientific, Singapore, 2010.


\bibitem{Mainardi-Gorenflo_FCAA2007}
F. Mainardi  and R. Gorenflo,  
\emph {Time-fractional derivatives in relaxation processes: a tutorial survey},
{Fract. Calc.  Appl. Anal.}, {\bf 10} (2007),   269--308. 
[E-print: {\tt arxiv.org/abs/0801.4914}]

\bibitem{Marichev_BOOK1983}
 O. I.  Marichev, 
 ``{Handbook of Integral Transforms of Higher Transcendental Functions,
 Theory and Algorithmic Tables}'',
 Ellis Horwood, Chichester, 1983..




 \bibitem{Mathai-Haubold_BOOK2008}
 A. M. Mathai  and H. J. Haubold,.
  ``{Special Functions for Applied Scientists}'',
 Springer, New York, 2008.

\bibitem{Mathai-Saxena_BOOK1978}
A. M. Mathai  and  R. K. Saxena,  
``{The H-Function with Applications in Statistics and Other Disciplines}'',
  Wiley Eastern Ltd, New Delhi,   1978. 


\bibitem{Mathai-Saxena-Haubold_BOOK2010}
 A. M. Mathai,  R. K. Saxena. and H. J. Haubold,  
 ``{The H-Function: Theory and Applications}'',
 Springer Verlag, New York,    2010.


\bibitem{Miller-Samko_ITSF2001}
 K. S. Miller and  S. G. Samko,   
\emph{Completely monotonic functions},
 {Integral Transforms and Special  Functions}, {\bf 12} (2001), 389--402.

\bibitem{Podlubny_BOOK1999}
I. Podlubny,  
``{Fractional Differential Equations}'',
Academic Press, San Diego, 1999.


\bibitem{Podlubny_MATLAB2006}
I. Podlubny, 
\emph{Mittag-Leffler function},  
 Matlab-Code that calculates the Mittag-Leffler function with desired accuracy,
  Matlab File Exchange
{\tt www.mathworks.com/matlabcentral/fileexchange}, 2006.

\bibitem{Pollard_BAMS1948}
        H. Pollard,
    \emph{The completely monotonic character of the Mittag-Leffler function}
     $E_\alpha (-x)$,
   { Bull. Amer. Math. Soc.},  {\bf 54} (1948), 1115--1116.



\bibitem{SKM_BOOK1993}
S. G. Samko, A. A.  Kilbas   and  O. I. Marichev,  
  ``{Fractional Integrals and Derivatives, Theory and Applications}'',
Gordon and Breach, Amsterdam, 1993..
   [English translation and revised version from the Russian edition,
``{Integrals and Derivatives of Fractional  Order and Some of  Their Applications}''
 Nauka i Tekhnika, Minsk, 1987]

\bibitem{Sandev_FLE_FCAA2012}
T. Sandev,  R. Metzler  and  Z. Tomovski,  
\emph{Velocity and displacement correlation functions for fractional generalized Langevin equations}, 
{Fract. Calc. Appl. Anal.}, {\bf 15} No 3 (2012), 426--450.
(DOI: 10.2478/s13540-012-0031-2)


\bibitem{Sansone-Gerretsen_BOOK1960}
  G. Sansone  and J. Gerretsen,  
``{Lectures on the Theory of Functions of a Complex Variable}'',
 Vol. I.  ``{Holomorphic Functions}'',
 Nordhoff, Groningen, 1960.  


\bibitem{SSV_BOOK2012}
R. L. Schilling,  R.  Song  and  Z. Vondra\v cek,   
``{Bernstein Functions. Theory and Applications}'',
2-nd ed., De Gruyter, Berlin, 2012.

\bibitem{Simon_2013}
T. Simon,
\emph{Comparing Fr\'echet and positive stable laws},
{Electron. J. Probab.}, {\bf 19} (2014), 1--25.
[E-print: {\tt arXiv:1310.1888} [math.PR] (2013),  pp. 27]


  \bibitem{Srivastava-Gupta-Goyal_BOOK1982}
H. M. Srivastava,  K. C.  Gupta  and  S. P. Goyal,  
``{The H-Functions of One and Two Variables with Applications}'',
South Asian Publishers, New Delhi and Madras, 1982.


\bibitem{Staroitov-Staroitova_SB2007}
A. P. Starovoıtov and N. A. Starovoıtova,
\emph{Pad\`e approximants of the Mittag-Leffler functions},
{Sbornik Mathematics},  {\bf 198} No 7 (2007), 1011--1023.
  (DOI: 10.1070/SM2007v198n07ABEH003871)

\bibitem{Tarasov_BOOK2011}
V. E. Tarasov, 
``Fractional Dynamics: Applications of Fractional Calculus to Dynamics of Particles, Fields and Media'',  
Springer, Berlin,  2011. 

\bibitem{Tomovski-et-al_ITSF2010}
Z. Tomovski,  R.  Hilfer . and H. M. Srivastava,   
\emph{Fractional and operational calculus with generalized fractional
derivative operators and Mittag-Leffler  type functions},
{Integral Transforms and Special  Functions}, {\bf 21} (2010),  797--814. 

\bibitem{Uchaikin_BOOK2013}
V. V. Uchaikin,
``Fractional Derivatives for Physicists and Engineers'',
Springer, Berlin, 2013.

\bibitem{Wong-Zhao_CA2002}
 R. Wong  and Y.-Q Zhao,  
\emph{Exponential asymptotics of the Mittag-Leffler function},
{Constructive Approximation}, {\bf 18} (2002), 355--385.

\bibitem{Zeng-Chen_PRE-PRINT2013}
C. Zeng  and Y.-Q. Chen,
\emph{Global Pad\`e approximations for the generalized Mittag-Leffler function and its inverse}, preprint,
 {\tt arXiv:1310.559} [math.CA] (2013),  pp. 6 (2 columns).

.

\end{thebibliography}
